\documentclass{article}

    \usepackage[final]{neurips_2024}

\usepackage[utf8]{inputenc} 
\usepackage[T1]{fontenc}    
\usepackage{hyperref}       
\usepackage{url}            
\usepackage{booktabs}       
\usepackage{amsfonts}       
\usepackage{nicefrac}       
\usepackage{microtype}      
\usepackage{xcolor}         
\usepackage{tabularx}

\usepackage{graphicx}
\usepackage{xcolor}
\definecolor{cobalt}{rgb}{0.03, 0.27, 0.49}
\definecolor{azurecolorwheel}{rgb}{0.0, 0.5, 1.0}
\definecolor{cadmiumgreen}{rgb}{0.0, 0.42, 0.24}
\usepackage{subcaption}
\usepackage{algorithm}
\usepackage{algpseudocode}
\usepackage{tikz-cd}
\usepackage{multirow}
\usepackage{amsmath}
\usepackage{wrapfig}

\title{Towards Secure and Private AI: A Framework for Decentralized Inference}

\author{%
  Hongyang Zhang, Yue Zhao, Claudio Angione, Harry Yang, James Buban,
  Ahmad Farhan, \\
  \textbf{Fielding Johnston}, \textbf{Patrick Colangelo}\\
  Nesa Research\\
  \texttt{research@nesa.ai}
}

\begin{document}

\maketitle

\begin{abstract}
The rapid advancement of ML models in critical sectors such as healthcare, finance, and security has intensified the need for robust data security, model integrity, and reliable outputs. Large multimodal foundational models, while crucial for complex tasks, present challenges in scalability, reliability, and potential misuse. Decentralized systems offer a solution by distributing workload and mitigating central points of failure, but they introduce risks of unauthorized access to sensitive data across nodes. We address these challenges with a comprehensive framework designed for responsible AI development. Our approach incorporates: 1) Zero-knowledge proofs for secure model verification, enhancing trust without compromising privacy. 2) Consensus-based verification checks to ensure consistent outputs across nodes, mitigating hallucinations and maintaining model integrity. 3) Split Learning techniques that segment models across different nodes, preserving data privacy by preventing full data access at any point. 4) Hardware-based security through trusted execution environments (TEEs) to protect data and computations. This framework aims to enhance security and privacy and improve the reliability and fairness of multimodal AI systems. Promoting efficient resource utilization contributes to more sustainable AI development. Our state-of-the-art proofs and principles demonstrate the framework's effectiveness in responsibly democratizing artificial intelligence, offering a promising approach for building secure and private foundational models.
\end{abstract}
\vspace{-0.2in}

\section{Introduction}
Artificial Intelligence (AI), particularly machine learning (ML), has made significant strides in recent decades, with multimodal systems integrating language, vision, and audio capabilities becoming increasingly prominent. 
The advent of large language models (LLMs) such as ChatGPT \cite{openai2023chatgpt}, Claude \cite{anthropic2023claude}, Gemini \cite{google2023gemini}, LLaMA \cite{llama2024} and diffusion models \cite{yang2023diffusion} such as DALLE-3 \cite{openai2023dalle3} and Sora \cite{deepmind2023sora} has ushered in a new era of foundation models with remarkable capabilities. While these multimodal foundation models exhibit intriguing properties such as in-context learning and chain-of-thought reasoning, they also present significant challenges in terms of security, privacy, reliability, and responsible deployment, especially in distributed or decentralized computing scenarios. 

These challenges are particularly acute when considering the potential for misuse, the generation of harmful content, and the propagation of misinformation. For instance, in ML as a service (MLaaS), ensuring the integrity and reliability of inference results is paramount. Service providers must demonstrate to customers that the output stems from a verified large language model, like GPT-4, rather than from human writers or less advanced models. This requirement, referred to as \textit{model security} or model integrity, is crucial for maintaining trust in AI systems. In distributed or decentralized inference applications, where a foundation model is divided into distinct segments managed by different parties, it is essential to verify the trustworthiness of each party's execution before aggregating outputs. This distributed approach, while offering advantages in scalability and robustness, introduces new challenges in ensuring consistent and reliable outputs across nodes.

Moreover, \textit{data privacy} issues are paramount in decentralized inference \cite{sun2019relationship}, particularly in ``critical inference'' settings involving sensitive information such as medical records, financial data, or security information. For example, in healthcare applications where AI models analyze medical images for disease diagnosis, protecting patient data from unauthorized access is crucial.

We aim to address these interleaved concerns—encompassing \textit{security} and \textit{privacy}—for the effective deployment of AI across various domains. Our approach involves designing hybrid solutions \cite{ma2022trusted} that can be tailored to specific use cases and security requirements. We discuss two key scenarios and our corresponding approaches, as well as how we adapt hardware-based trusted execution environments (TEEs) as an orthogonal approach to enhance security and privacy.

\textbf{Scenario 1: Critical inference}
refers to scenarios where the results of AI inference are extremely significant, necessitating the highest levels of security and privacy, even if it means slower speed and higher cost. 
In these situations, the accuracy and confidentiality of the inference outcomes are important, and users are willing to endure longer processing times to ensure their data is fully protected. 
Examples include healthcare diagnostics, where AI analyzes medical images to detect diseases, and financial decision-making, where AI evaluates large transactions or investment strategies. 
In both cases, the sensitive nature of the data demands robust privacy measures, with stakeholders prioritizing data security over rapid results to prevent unauthorized access and ensure the integrity of the process. Under this scenario, we adapt and innovate zero-knowledge machine learning (ZKML) \cite{sun2023zkdl} for model integrity verification. 
Admittedly, the original design of ZKML is computationally heavy, especially for non-linear layers in neural networks such as ReLU layers \cite{liu2021zkcnn}. 
To balance effectiveness and efficiency, we apply the state-of-the-art ZKML techniques introduced by \cite{sun2023zkdl,sun2024zkllm} to build \textit{zero-knowledge Decentralized Proof System (zkDPS)} for proof generation and verification processes in foundation models.
\S \ref{sec:critical-inference} provides more details on our customized zkDPS and SHE solutions for critical inference with the highest protection.

\textbf{Scenario 2: General inference}
refers to everyday AI inference tasks with less critical results, allowing for faster processing speeds and lower costs without compromising basic security and privacy standards. 
In these scenarios, a certain level of protection is still necessary, but the stringent measures required for critical inference are not needed. Examples include routine tasks such as checking the weather, recommending products, or filtering spam emails. 
Here, the emphasis is on efficiency and speed while maintaining adequate data security to protect user information. 
Under this scenario, we employ our \textit{consensus-based verification (CBV)} for security, ensuring the integrity of the inference process, and \textit{split learning (SL)} \cite{vepakomma2018split} for data encryption, which provides a reasonable level of data protection.
Notably, we have innovated this verification method to reduce the high redundancy requirements in model verification, and split learning has been used in decentralized AI encryption for the first time. 
\S \ref{sec:general-inference} elaborates on our consensus-based verification and split learning solutions for general inference, balancing speed and security effectively.

\textbf{Our Trusted Execution Environment (TEE)}.
In addition to our innovations around software- and/or algorithm-based security and privacy approaches, we also design an orthogonal hardware-based approach. 
In a nutshell, TEEs create secure isolation zones within the network's nodes, protecting user data and private model parameters from unauthorized access \cite{sabt2015trusted}. TEEs provide a secure enclave for executing computations, isolated from the rest of the node's operating environment, ensuring that even if other parts of the node are compromised, the computations within the TEE remain protected. 
This isolation is critical in decentralized settings where AI models are spread among multiple owners. 
This hardware-based security measure works as an alternative to algorithm-based approaches,
providing a fast and efficient solution for our decentralized AI inference tasks. 
Additionally, we innovate our TEEs by optimizing communication among multiple TEEs by establishing direct, secure channels, and implementing heterogeneous TEE scheduling based on the capabilities of each node, whether CPU or GPU-based. 
Our TEE facilitates secure collaboration across multiple nodes, enabling us to maintain high performance, robust security, and data privacy, making it ideal for both critical and general inference scenarios. 
See details of our TEE implementation and innovation in \S \ref{sec:tee}.

\begin{table}[ht]
\centering
\small
\caption{Summary of our proposed security and privacy solutions.}
\label{table:security-all}
\begin{tabular}{|p{1.2cm}|p{1.6cm}|p{4cm}|p{5cm}|}
\hline
\textbf{Type} & \textbf{Scenario} & \textbf{Solution to Model Verification} & \textbf{Solution to User Privacy} \\
\hline
Algorithm & Critical (\S\ref{sec:critical-inference}) & zero-knowledge Decentralized Proof System (zkDPS) (\S\ref{subsec:zkp}) & Sequential Homomorphic Encryption (SHE) (proprietary)\\
\hline
Algorithm & General (\S\ref{sec:general-inference}) & Consensus-Based Verification (CBV) (\S\ref{subsec:cbv}) & Split Learning (SL) (\S\ref{subsec:sl}) \\
\hline
Hardware & Both & TEE model verification (\S\ref{sec:tee}) & TEE secure channel (\S\ref{sec:tee}) \\
\hline
\end{tabular}
\end{table}

\textbf{Choosing Security and Privacy Options}.
Table \ref{table:security-all} summarizes our innovations around security and privacy. Here, we further discuss the choice of different methods by scenario. Given critical inference scenarios, where the final results have high value and users are willing to wait longer, we suggest using zkDPS for model verification and SHE for data encryption. In contrast, for general inference scenarios, where the results are less critical and faster processing is desired, we recommend using CBV for model verification and SL for data encryption. Additionally, the TEE-based solution, our specialized TEE, can also be leveraged as an orthogonal approach. Once our TEE is employed, it addresses both verification and encryption needs simultaneously. However, it is important to note that GPU TEEs are only available on high-end GPUs, making the trade-off such that TEE solutions are mostly CPU-based. In contrast, algorithm-based solutions can fully leverage GPUs, offering a different balance of performance and security. This approach allows us to balance the trade-offs between security, privacy, and efficiency, ensuring that users can select the most appropriate solution for their specific needs. By leveraging both algorithm-based and hardware-based security measures, we can provide robust and adaptable security and privacy solutions across a wide range of AI applications. This adaptive approach ensures that users can select the most appropriate level of protection for their needs, achieving an optimal balance among security, privacy, and performance.

\section{Security and Privacy for Critical Inference}
\label{sec:critical-inference}

\begin{wrapfigure}{r}{0.6\textwidth}
    \centering
    \includegraphics[width=\linewidth]{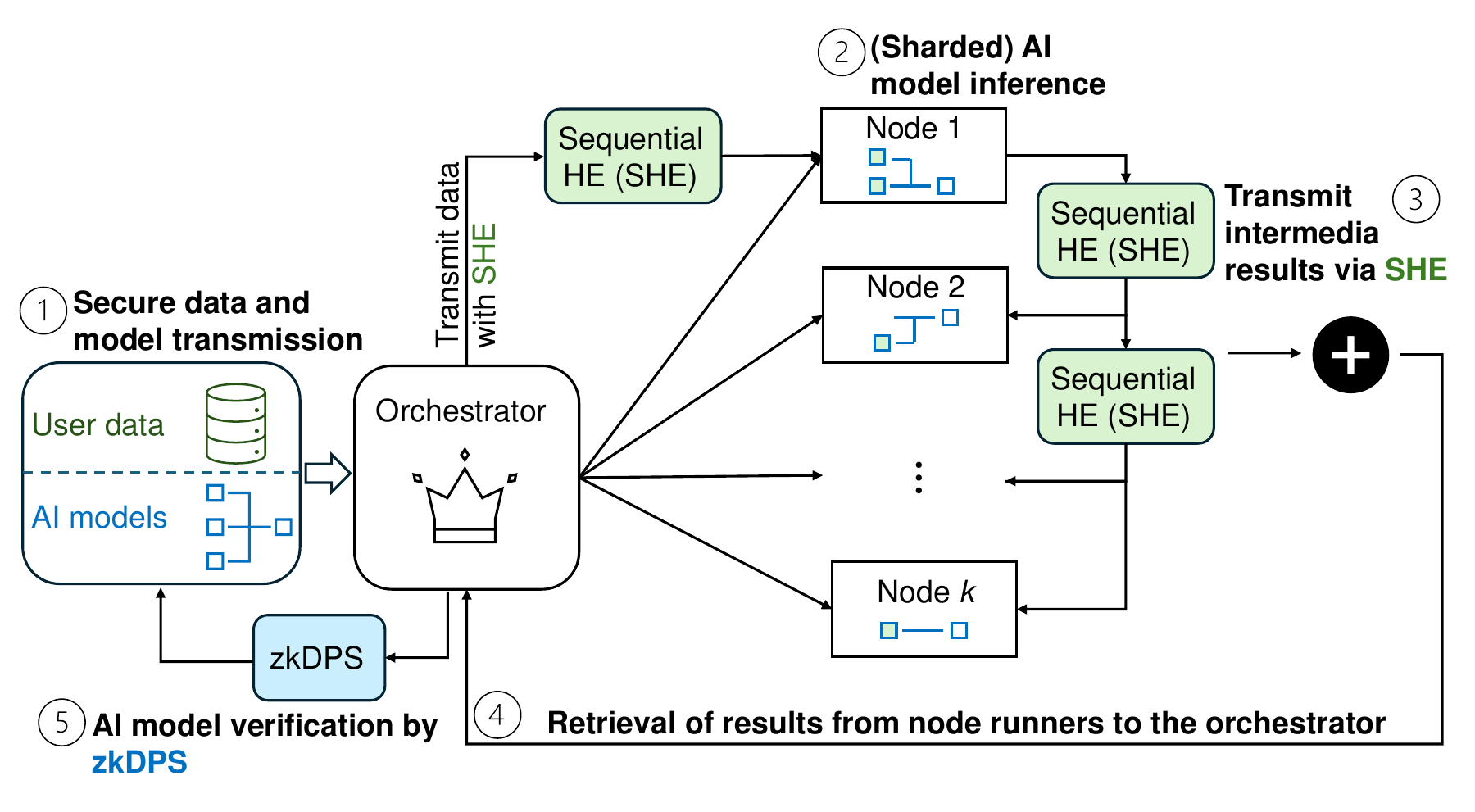}
    \caption{Security flowchart for critical inference, where we design \textcolor{azurecolorwheel}{zkDPS} (\S \ref{subsec:zkp}) for model verification and \textcolor{cadmiumgreen}{SHE} (proprietary) for data encryption and user privacy protection.}
    \label{fig:critical-flow}
\end{wrapfigure}

In critical inference scenarios, the significance of AI inference results necessitates the highest levels of security and privacy, as the outcomes directly impact crucial sectors like healthcare and finance. 
Our strategy for ensuring model verification and integrity is via our Zero-Knowledge Decentralized Proof System (zkDPS), a specialized ZK system for decentralized LLMs that allows one party to prove to another that a statement is true without revealing any information beyond the validity of the statement itself. Notably, it introduces a few new techniques to speed up the ZK process.
This approach is detailed in \S \ref{subsec:zkp}.

Fig. \ref{fig:critical-flow} summarizes the security flow of critical inference, where the data will be transmitted with SHE for encryption, and the final inference results will be verified by zkDPS for integrity.
Due to the space limitation, the extensive technical details are deferred to Appendix \ref{sec:critical-inference-appx}.

\newpage
\section{Security and Privacy for General Inference}
\label{sec:general-inference}

\begin{wrapfigure}{r}{0.6\textwidth}
    \centering
    \includegraphics[width=\linewidth]{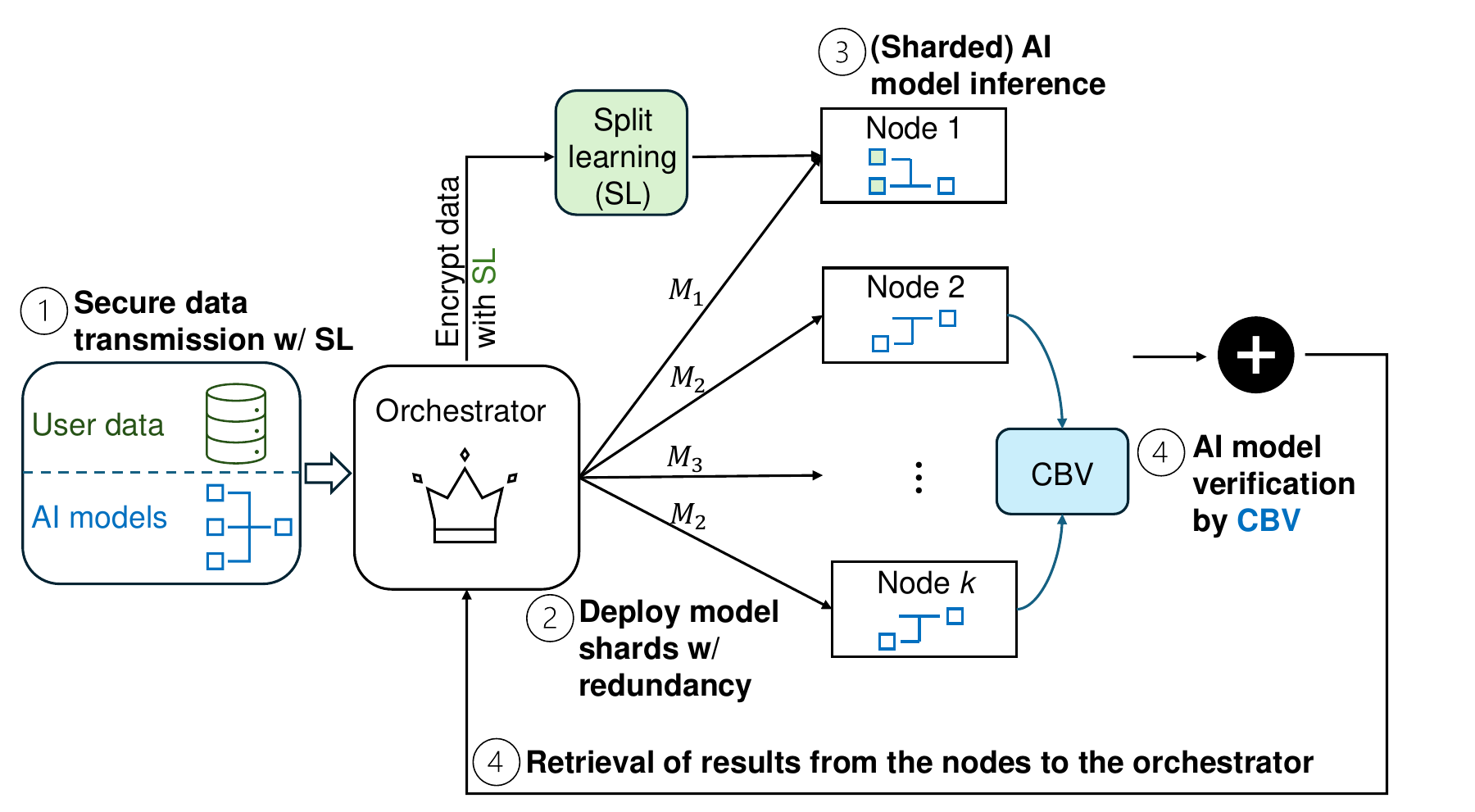}
    \caption{Security flowchart for general inference, where we design \textcolor{azurecolorwheel}{CBV} (\S \ref{subsec:cbv}) for model verification and leverage \textcolor{cadmiumgreen}{SL} (\S \ref{subsec:sl}) for data encryption and user privacy protection.}
    \label{fig:general-flow}
\end{wrapfigure}
In general inference scenarios, the results of AI inference tasks are less critical, allowing for faster processing speeds and lower costs without compromising basic security and privacy standards. These tasks include everyday activities such as checking the weather, recommending products, or filtering spam emails. Our strategy for ensuring model verification and integrity in these scenarios is Consensus-based Verification Check (CBV), which leverages the collective agreement of multiple nodes to ensure the correctness and integrity of model execution without revealing sensitive data. This approach is detailed in Section \ref{subsec:cbv}. 

For data privacy, we employ split learning (SL) \cite{vepakomma2018split,thapa2022splitfed}, a method where the model is divided into segments, and each segment is trained on different nodes to maintain data privacy by ensuring no single node has access to the complete dataset. More information can be found in Section \ref{subsec:sl}.

Fig. \ref{fig:general-flow} provides an overview of the security flow of general inference in our system, which first applies SL to encrypt the first (few) layers to ``encrypt'' the raw data, where the AI models are sharded and deployed to nodes with redundancy. The outputs of the nodes who have the same shard will be verified by consensus for integrity.
Due to the space limitation, the extensive technical details are deferred to Appendix \ref{sec:general-inference-appx}.

\section{Conclusions and Future Directions}

Artificial intelligence and machine learning have made remarkable advancements, particularly in multimodal systems integrating language, vision, and audio capabilities. While foundation models like ChatGPT, DALLE-3, and Sora demonstrate impressive capabilities, they also raise significant concerns regarding security, privacy, reliability, and responsible deployment, especially in distributed and decentralized computing scenarios. Our work addresses these challenges by offering adaptive solutions tailored to different scenarios, considering both technical and ethical implications.

We propose a comprehensive framework that ensures inference integrity and data privacy while contributing to the reliability and fairness of multimodal AI systems. For critical tasks, we recommend a zero-knowledge Decentralized Proof System (zkDPS) and Sequential Homomorphic Encryption (SHE). For general tasks prioritizing efficiency, we propose Consensus-based Verification (CBV) and Split Learning (SL). Additionally, our Trusted Execution Environment (TEE) offers a hardware-based solution addressing both verification and encryption needs. This hybrid strategy balances security, privacy, and responsible AI development, providing users with flexible, scenario-specific solutions.

Looking ahead, we aim to optimize zkDPS for real-time applications, develop adaptable ZK templates for diverse models, and create an automated framework for recommending optimal security approaches. We also plan to investigate methods to enhance the reliability of multimodal models, improve fairness, reduce bias, and develop sustainable practices for AI deployment. These efforts will focus on addressing hallucinations, misinformation propagation, and ethical considerations in decentralized systems.

By pursuing these directions, we strive to improve the robustness, efficiency, and adaptability of our solutions while contributing to the responsible development of next-generation multimodal foundational models. Our work represents a significant step towards building more secure, reliable, and ethically sound AI systems, integrating security and privacy considerations with broader responsible AI principles to foster technologies that are not only powerful and efficient but also trustworthy.

\clearpage
\newpage

\bibliographystyle{unsrt}
\bibliography{references}

\newpage
\clearpage

\appendix
\section*{Supplementary Material}

\setcounter{table}{0}
\setcounter{figure}{0}
\setcounter{algorithm}{0}

\section{Details of Security and Privacy for Critical Inference}
\label{sec:critical-inference-appx}

In critical inference scenarios, the significance of AI inference results necessitates the highest levels of security and privacy, as the outcomes directly impact crucial sectors like healthcare and finance. 
Our strategy for ensuring model verification and integrity is via our Zero-Knowledge Decentralized Proof System (zkDPS), a specialized ZK system for decentralized LLMs that allows one party to prove to another that a statement is true without revealing any information beyond the validity of the statement itself. Notably, it introduces a few new techniques to speed up the ZK process.
This approach is detailed in \S \ref{subsec:zkp}. 
Fig. \ref{fig:critical-flow} summarizes the security flow of critical inference, where the data will be transmitted with SVE for encryption, and the final inference results will be verified by zkDPS for integrity.

\subsection{Zero-Knowledge Machine Learning for Model Integrity}

\label{subsec:zkp}
\subsubsection{Background}

Despite the significant strides made in AI security in recent years, the potency of attacks has surged. Central to AI security is the pivotal task of delineating the threat model and understanding how adversaries target the inference process. Adversaries can exploit various vulnerabilities within the inference systems of foundational models, employing tactics tailored to different scenarios. In decentralized AI inference environments, one threat model emerges, where computing nodes may behave deceitfully, compromising the integrity of aggregated results. It becomes imperative to establish mechanisms wherein each node can verify its adherence to agreed-upon protocols without compromising the confidentiality of its model. This necessitates enabling nodes to provide proofs of honest execution to the central server or the public while safeguarding the confidentiality of their respective models. Thus, ensuring both the integrity of inference processes and the privacy of model architectures becomes paramount in the realm of AI security.

In the realm of defending against adversarial attacks, a plethora of meticulously crafted countermeasures exist to safeguard systems. One such strategy, particularly pertinent in decentralized inference settings, involves the implementation of mechanisms where central servers solicit proof of computing execution from each node. This process is orchestrated with precision to ensure that while the proof is furnished, the node's confidential model parameters remain undisclosed. Subsequently, the central servers meticulously scrutinize the proofs submitted, thereby enabling them to discern the reliability of each node. The efficacy of this approach hinges upon the successful verification of the proof, serving as a litmus test for the trustworthiness of the node in question.


The challenge intensifies when fast inference of secure foundation models is required. Given the scale of big data and the models involved, foundation models inherently exhibit slow inference speeds. It is widely acknowledged that incorporating security measures further exacerbates this slowdown in AI models. For instance, the fastest-known zero-knowledge proof algorithm currently takes as long as 15 minutes to generate proof for a single token in the output~\cite{sun2024zkllm}. Despite substantial efforts to expedite the inference process, it is imperative to ensure the security of foundation model inference without compromising efficiency.

\subsubsection{Problem Setups}

\textbf{Decentralized Inference.}
We explore the challenge of decentralized inference for foundation models, which offers numerous benefits. With the advent of 5G technology and improved internet latency, personalized devices such as mobile phones can now participate in crowdsourcing machine learning models. This approach enhances device utilization and eliminates the need for data communication between nodes and a central server, thereby safeguarding data privacy. In the context of Machine Learning as a Service (MLaaS), models remain the private property of individual nodes, allowing them to offer model inference services via APIs without sharing model weights and checkpoints. As foundation models grow in size—such as Meta’s LLaMA-3.1 with 405 billion parameters~\cite{llama2024}, and future models expected to reach trillion-level sizes—it becomes impractical to load entire models on a single node or server. Additionally, technologies like blockchain are now equipped to support the decentralized inference of foundation models on-chain, further facilitating this approach.

\medskip
\textbf{Model Decomposition.}
A core assumption in decentralized AI inference is that the model used in the inference system can be divided into several parts, each managed by a separate party, or computing node. Each computing node performs its assigned computations and sends the results to a central server. For instance, with a foundation model like LLaMA-3, the model can be decomposed layer-wise, enabling sequential inference by different nodes. Alternatively, decomposing the model width-wise allows for parallel inference across multiple nodes. The method of decomposition depends entirely on the application scenarios, and we consider both approaches in our products.

\subsubsection{Threat Models}

In this paper, we focus on decentralized inference of foundation models as described above. In this framework, each computing node (i.e. prover) owns a foundation model or a part of the foundation model with a publicly known architecture, while the model weights are proprietary. We make a semi-dishonest assumption on the central server (i.e. verifier): the central server is honest in aggregating results from each computing node and accurately reports the verification result, but the central server tries to glean additional information about parts of the foundation model at computing nodes. However, some computing nodes might be dishonest, potentially deviating from the agreed protocol by substituting their model with an alternative or outputting random data. We assume that the majority of nodes are honest, but acknowledge that dishonest nodes can collude. These adversarial nodes can arbitrarily alter their results, provided their behavior remains undetected by the central server.

\subsubsection{Overview of Zero-Knowledge Proofs}
\label{sec:zkp}

\begin{figure}
    \centering
    \includegraphics[width=0.85\textwidth]{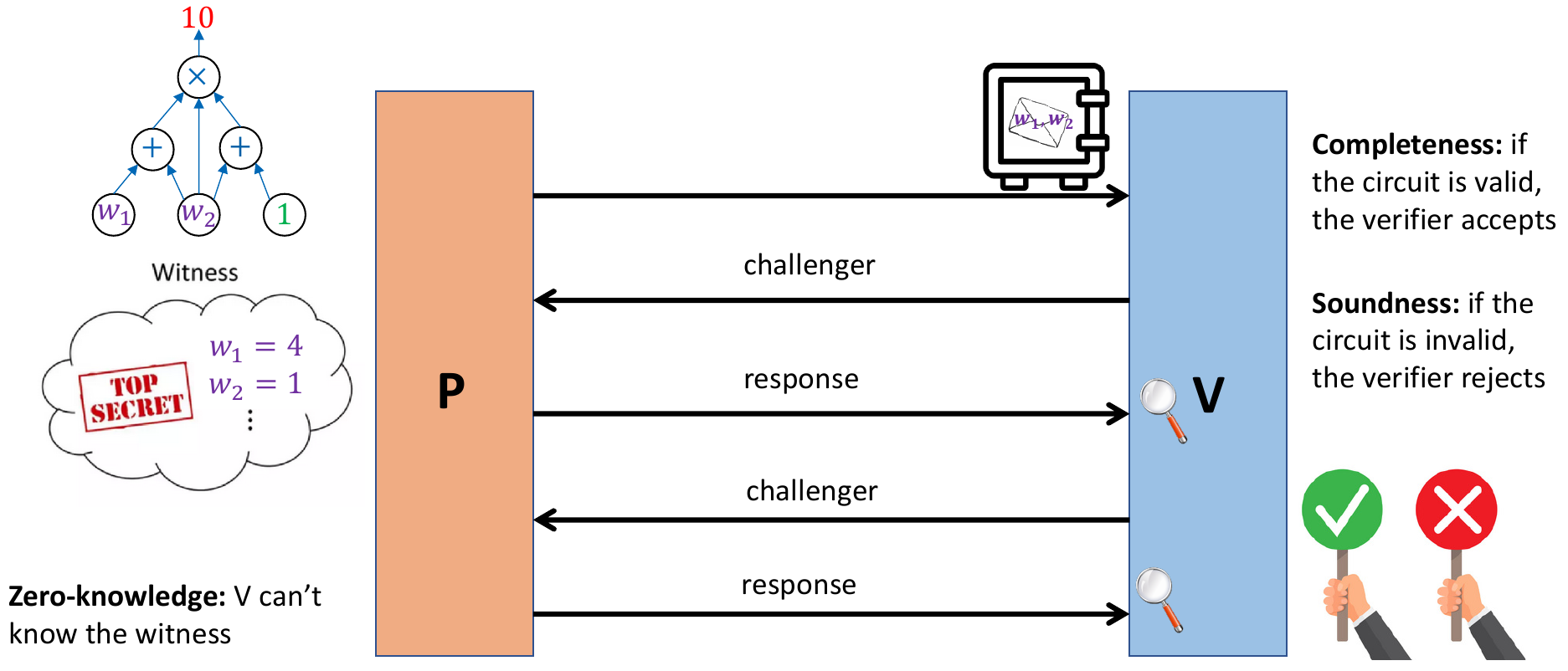}
    \caption{Zero-knowledge proof of circuit $10=(w_1+w_2)(w_2+1)$ between a prover (P) and a verifier (V). Hereby, the goal of the prover is to prove to the verifier that P knows a $w_1$ and $w_2$ such that the claimed result ``10'' is indeed calculated by the equation $(w_1+w_2)(w_2+1)$ (which is denoted by a circuit). The witness $w_1=4$ and $w_2=1$ are the secret of the prover. Zero-knowledge proof consists of a commitment process (denoted by the safe box) in the beginning, followed by several back-and-forth challenge and response processes between P and V in the interactive scenario. In the non-interactive scenario, the prover can challenge him or herself by the Fiat-Shamir heuristic and the verifier only needs to verify the last response from the prover.}
    \label{fig:zkp}
\end{figure}

We use the technique of zero-knowledge proofs to guarantee that each party is honest about his or her execution of inference of foundation models. Zero-knowledge proof serves as a fundamental technique and underpins the architecture of blockchain. In this cryptographic concept, two entities are involved: the prover and the verifier. The prover's objective is to demonstrate the successful execution of a protocol without disclosing confidential information, termed as the 'witness'. This witness encompasses sensitive data like model weights or private information that the prover wishes to keep undisclosed to the verifier. Often, the protocol is depicted as a circuit, where certain components remain hidden within the witness.

\medskip
\textbf{Commitment, Proof, and Verification.}
The process of zero-knowledge proof involves three essential steps. Firstly, the prover commits to the witness data, such as model parameters, ensuring its integrity by encrypting it before transmitting it to the verifier. Once sent, the content remains unchanged, with the verifier gaining access only to the encrypted version. In practice, the (generalized Pedersen) commitment of any $d$-dimensional tensor $\textbf{S}=(S_1,S_1,...,S_d)\in\{0,1,...,|\mathbb{G}|-1\}^d$ (e.g., the model weights) is implemented as \cite{wahby2018doubly}:
{\small
\begin{equation}
\label{equ: Pedersen}
    \textsf{Commit}(\textbf{S},r_S)=h^{r_S}\textbf{g}^\textbf{S}=h^{r_S}\prod_{i=1}^{d} g_i^{S_i},
\end{equation}
}
where $\mathbb{G}$ is an elliptic curve group w.r.t. the finite field $\mathbb{F}$ consisting of points $(x,y)\in\mathbb{F}\times \mathbb{F}$ such that $y^2=x^3+ax+b$ for designated field elements $a$ and $b$, $r\sim \{0,1,...,|\mathbb{G}|-1\}$ is an integer (i.e., the element of the scalar field $\mathbb{F}$ of $\mathbb{G}$ with $|\mathbb{F}|=|\mathbb{G}|$) uniformly sampled from $\{0,1,...,|\mathbb{G}|-1\}$, $\textbf{g}=(g_1,g_2,...,g_d)\in\mathbb{G}^d$ and $h\sim\mathbb{G}$ are uniformly and independently sampled from $\mathbb{G}$. In the second step, the prover executes the protocol and simultaneously generates a proof of execution in a finite field $\mathbb{F}$, which is then transmitted to the verifier, either interactively or non-interactively. Depending on the protocol, operations can involve arithmetic or non-arithmetic processes. Lastly, the verifier meticulously examines the proof to ensure the honest execution of the protocol by the prover, thereby validating the transaction or information under scrutiny.

\subsubsection{Properties of Zero-Knowledge Proofs}
Zero-knowledge proofs have many advantageous properties that form the foundation of blockchain and decentralized machine learning. These include:

\textbf{Completeness:} If the prover accurately executes the circuit, the proof will be validated (with probability 1).

\textbf{Special Soundness:} If the prover is dishonest in executing the circuit, the proof will fail (with high probability). A weaker property, \emph{special soundness}, requires executing the protocol twice and being able to identify the witness.

\textbf{Zero-Knowledge:} The verifier gains no knowledge about the prover's witness.

The above properties ensure that the proof will pass verification only if the prover is honest, while also allowing the prover to keep its secret or witness hidden from the verifier.

\medskip
\textbf{Interactive vs. Non-Interactive Proof Systems.}
Depending on whether the proof-verification process involves a single round or multiple rounds, zero-knowledge proofs can be classified as either interactive or non-interactive, respectively. Zero-knowledge proofs are naturally described as an interactive process, where the verifier sends a challenge (typically a random variable) to the prover, who then responds to the verifier. If the proof is valid, the verifier sends a new challenge in the next round, and the process repeats (see Figure \ref{fig:zkp}). Given an input $x$, an interactive proof procedure works as follows:
\begin{enumerate}
    \item $P$ sends the first message $\alpha\leftarrow P(x)$.
    \item $V$ sends a challenge $\beta$.
    \item $P$ sends the second message $\gamma\leftarrow P(x,\alpha,\beta)$.
    \item $V$ decides to accept or reject according to an algorithm $V(x,\alpha,\beta,\gamma)$.
\end{enumerate}
In the non-interactive case, the process can be simulated by having the prover generate his/her own challenge $\beta$. This is achieved by replacing the random variable $\beta$ in the challenge with a random oracle model or a hash function $H$ (such as SHA-256) that operates on all the messages that the prover has sent so far. This is also known as the Fiat-Shamir heuristic~\cite{fiat1986prove}. In particular, given an input $x$, a one-shot, non-interactive proof procedure works as follows:
\begin{enumerate}
    \item $P$ computes the first message $\alpha\leftarrow P(x)$.
    \item $P$ computes a challenge $\beta\leftarrow H(x,\alpha)$.
    \item $P$ computes the second message $\gamma\leftarrow P(x,\alpha,\beta)$.
    \item $P$ sends $\alpha$ and $\gamma$ to $V$.
    \item $V$ computes $\bar{\beta}\leftarrow H(x,\alpha)$ and decides to accept or reject according to an algorithm $V(x,\alpha,\bar{\beta},\gamma)$.
\end{enumerate}

\subsubsection{Commitment}
The Pedersen commitment \eqref{equ: Pedersen} satisfies the binding property: once sent to the verifier, the opening information $(r,\mathbf{S})$ cannot be changed anymore by the prover. Upon initial inspection, the prover needs to send the witness $\mathbf{S}$ to the verifier to prove that he/she can ``open'' the commitment. Thus the zero-knowledge property of the commitment \eqref{equ: Pedersen} may appear impossible.
However, this skepticism is unfounded, largely owing to the homomorphic property of the Pedersen commitment \eqref{equ: Pedersen}: for two commitments $\textsf{Commit}(\textbf{S}_1,r_1)$ and $\textsf{Commit}(\textbf{S}_2,r_2)$ corresponding to tensors $\textbf{S}_1$ and $\textbf{S}_2$, respectively, we have $\textsf{Commit}(\textbf{S}_1,r_1)\cdot \textsf{Commit}(\textbf{S}_2,r_2)=\textsf{Commit}(\textbf{S}_1+\textbf{S}_2,r_1+r_2)$, responding to the commitment of tensor $\textbf{S}_1+\textbf{S}_2$. This property enables the prover to prove to the verifier that he/she can ``open'' the commitment without revealing the witness. This is achieved by letting the prover instead open the commitment of a linear transformation of the witness: $\mathbf{S}'\leftarrow\mathbf{S}\cdot e+\mathbf{D}$, where $\mathbf{D}$ is a $d$-dimensional hiding vector picked by the prover and $e$ is a challenge (scalar) randomly sampled by the verifier. Hereby, the vector $\mathbf{D}$ is to hide the witness $\mathbf{S}$ as $\mathbf{S}\cdot e+\mathbf{D}$ looks random to the verifier. The existence of challenge $e$ guarantees special soundness as two runs of the procedure with challenges $e_1$ and $e_2$ with $e_1\not=e_2$ satisfy:
\begin{equation}
\label{equ: two runs}
\begin{split}
    &\mathbf{S}_1'=\mathbf{S}\cdot e_1+\mathbf{D},\\
    &\mathbf{S}_2'=\mathbf{S}\cdot e_2+\mathbf{D}.\\
\end{split}
\end{equation}
In Equation \eqref{equ: two runs}, we have $2d$ unknowns $\mathbf{S}$ and $\mathbf{D}$ and $2d$ equations. Thus, two accepting transcripts $(\mathbf{S},e_1,\mathbf{D})$ and $(\mathbf{S},e_2,\mathbf{D})$ will identify $\mathbf{S}$ and $\mathbf{D}$ by Gaussian elimination, thus achieving special soundness.

\begin{algorithm}
\caption{Committer/Prover proves to a verifier that he/she knows a witness $\mathbf{S}\in\mathbb{F}^d$ and a witness $v\in\mathbb{F}$ which are openings of the commitments \eqref{equ: Pedersen} and \eqref{equ: t commitment}, respectively, such that $\langle\mathbf{S},\mathbf{y}\rangle=t$. Denote by $p=|\mathbb{F}|=|\mathbb{G}|$.}
\label{alg:commitment}
\begin{algorithmic}[1]
\State Randomly sample generators $h,g,g_1,...,g_d$ from the elliptic curve group $\mathbb{G}$, which are publicly known to both parties.
\State Prover sends the commitment $c_S=\textsf{Commit}(\textbf{S},r_S):=h^{r_S}\prod_{i=1}^{d} g_i^{S_i}$ and $c_t=\textsf{Commit}(t,r_t):=h^{r_t} g^t$ to verifier. Hereby, prover knows $\mathbf{S}$, $t$, $r_S$, $r_t$, $c_S$ and $c_t$. The verifier only knows $c_S$ and $c_t$.
\State Prover picks $\mathbf{D}\in\{0,1,...,p-1\}^d$ and hiding factors $r_1,r_2\in\{0,1,...,p\}$ and sends to verifier the commitment $c_D=\textsf{Commit}(\textbf{D},r_1):=h^{r_1}\prod_{i=1}^{d} g_i^{D_i}$ and $c_{\langle \mathbf{D},\mathbf{y}\rangle}=\textsf{Commit}(\langle \mathbf{D},\mathbf{y}\rangle,r_2):=h^{r_2} g^{\langle \mathbf{D},\mathbf{y}\rangle}$.
\State Verifier randomly picks a challenge $e\in\{0,1,...,p-1\}$ and sends it to prover.
\State Prover computes $\mathbf{S}':=\mathbf{S}\cdot e+\mathbf{D}$ and $r_{S'}:=r_S\cdot e+r_1$.
\State Similarly, prover computes $r_{t'}:=r_t\cdot e+r_2$.
\State Prover sends $(\mathbf{S}',r_{S'})$ and $r_{t'}$ to verifier. Note that $\mathbf{S}'$ and $r_{S'}$ reveal no information about $\mathbf{S}$ and $r_S$ to verifier, and $r_{t'}$ reveal no information about $t$ and $r_t$ to verifier.
\State Verifier computes  $c_{S'}=\textsf{Commit}(\textbf{S}',r_{S'}):=h^{r_{S'}}\prod_{i=1}^{d} g_i^{S_i'}$ and checks whether $c_{S'}\overset{\text{?}}{=}c_S^e\cdot c_D$, where the RHS is the expected commitment of $\mathbf{S}'$ from the perspective of verifier by additive homomorphism.
\State Verifier computes $t':=\langle\mathbf{S}',\mathbf{y}\rangle$ and $c_{t'}=\textsf{Commit}(t',r_{t'}):=h^{r_{t'}} g^{t'}$ and checks whether $c_{t'}\overset{\text{?}}{=}c_t^e\cdot c_{\langle \mathbf{D},y\rangle}$, where the RHS is the expected commitment of $t'$ from the perspective of verifier by additive homomorphism.
\end{algorithmic}
\end{algorithm}

Algorithm \ref{alg:commitment} provides a complete procedure for the opening of the commitment. Let $\mathbf{y}$ be a publicly known vector to both parties. The algorithm enables the prover to prove to the verifier that he/she knows a witness $\mathbf{S}\in\mathbb{F}^d$ and a witness $t\in\mathbb{F}$ which are openings of the commitment \eqref{equ: Pedersen} and the commitment of $t$:
\begin{equation}
\label{equ: t commitment}
\textsf{Commit}(t,r_t):=h^{r_t}g^t,
\end{equation}
such that $\langle\mathbf{S},\mathbf{y}\rangle=t$, where $h$ and $g$ are generators sampled randomly from an elliptic curve group. The algorithm is provably complete, has special soundness, and exhibits zero-knowledge properties~\cite{thaler2022proofs}. To see this, the completeness and zero-knowledge are obvious by the inspection of the algorithm. To see the special soundness, for two runs of the algorithm with challenges $e_1$ and $e_2$ ($e_1\not= e_2$) from the verifier, by Lines 8-9 of Algorithm \ref{alg:commitment}, we have
\begin{subequations}
\begin{align}
    &h^{r_{S_1'}}\prod_{i=1}^{d} g_i^{S_{1i}'}=c_S^{e_1} \cdot c_D, \label{equ:1}\\
    &h^{r_{S_2'}}\prod_{i=1}^{d} g_i^{S_{2i}'}=c_S^{e_2} \cdot c_D, \label{equ:2}\\
    &h^{r_{t_1'}} g^{\langle\mathbf{S}_1',\mathbf{y}\rangle}=c_t^{e_1}\cdot c_{\langle\mathbf{D},\mathbf{y}\rangle}, \label{equ:3}\\
    &h^{r_{t_2'}} g^{\langle\mathbf{S}_2',\mathbf{y}\rangle}=c_t^{e_1}\cdot c_{\langle\mathbf{D},\mathbf{y}\rangle},\label{equ:4}
\end{align}
\end{subequations}
where $S_{1i}'$ and $S_{2i}'$ represent the $i$-th element of vector $\mathbf{S}_1'$ and $\mathbf{S}_2'$, respectively.
Denote by
\begin{equation}
\begin{split}
    & r_{\bar{S}} := (r_{S_1'}-r_{S_2'})/(e_1-e_2) \text{ mod } |\mathbb{G}|,\\
    & r_{\bar{t}} := (r_{t_1'}-r_{t_2'})/(e_1-e_2) \text{ mod } |\mathbb{G}|,\\
    & \bar{\mathbf{S}} := (\mathbf{S}_1'-\mathbf{S}_2')/(e_1-e_2) \text{ mod } |\mathbb{G}|.
\end{split}
\end{equation}
Dividing Equation \eqref{equ:1} by \eqref{equ:2}, we have
\begin{equation}
    h^{r_{\bar{S}}}\prod_{i=1}^d g_i^{\bar{S}_i}=c_S.
\end{equation}
Similarly, dividing Equation \eqref{equ:3} by \eqref{equ:4}, we have
\begin{equation}
    h^{r_{\bar{t}}} g^{\langle\bar{\mathbf{S}},\mathbf{y}\rangle}=c_t.
\end{equation}
Thus, we have shown that $(\bar{\mathbf{S}},r_{\bar{S}})$ is an opening of the commitment $c_S$ and $(\langle\bar{\mathbf{S}},\mathbf{y}\rangle,r_{\bar{t}})$ is an opening of the commitment $c_t$, as desired.


\subsubsection{Proofs for Arithmetic Operations}

\medskip
\textbf{Multilinear Extension.} Zero-knowledge proof focuses on the finite field $\mathbb{F}$, rather than the real field $\mathbb{R}$ as in the floating-point calculations. Arithmetic operations consist of addition and multiplication. For these arithmetic operations, it is easy to use the Sum-Check protocol, in particular, the GKR protocol~\cite{goldwasser2015delegating,thaler2015note}, to implement zero-knowledge proofs. The basic idea in the Sum-Check protocol is to express a $d$-dimensional tensor $\mathbf{S}\in\mathbb{F}^d$ involved in the calculation as a multi-variable polynomial $\widetilde{\mathbf{S}}:\mathbb{F}^{\log_2 d}\rightarrow \mathbb{F}$ via a transformation called multilinear extension~\cite{thaler2022proofs}:
\begin{equation}
    \widetilde{\mathbf{S}}(\mathbf{u})=\sum_{\mathbf{b}\in\{0,1\}^{\log_2 d}} \mathbf{S}(\mathbf{b})\widetilde\beta(\mathbf{u},\mathbf{b}),
\end{equation}
where $\mathbf{b}\in\{0,1\}^{\log_2 d}$ refers to the binary index of tensor $\mathbf{S}$, and $\widetilde\beta(\cdot,\cdot):\mathbb{F}^{\log_2 d}\times \mathbb{F}^{\log_2 d}\rightarrow \mathbb{F}$ is the (unique) Lagrangian interpolation polynomial:
\begin{equation}
    \widetilde\beta(\mathbf{u},\mathbf{b})=\prod_{i=1}^{\log_2 d} (u_ib_i+(1-u_i)(1-b_i)),
\end{equation}
 such that for any $\mathbf{b}_1,\mathbf{b}_2\in\{0,1\}^{\log_2 d}$, the interpolation property holds true:
\begin{equation}
    \widetilde\beta(\mathbf{b}_1,\mathbf{b}_2)=
    \begin{cases}
        1, & \text{if }\mathbf{b}_1=\mathbf{b}_2;\\
        0, & \text{otherwise}.
    \end{cases}
\end{equation}
That is, the polynomial $\widetilde{\textbf{S}}$ is the Lagrange interpolation of the tensor $\textbf{S}$ on the binary indices: $\widetilde{\textbf{S}}$ is the unique multilinear polynomial over $\mathbb{F}$ such that $\widetilde{\textbf{S}}(\mathbf{u})=\textbf{S}(\mathbf{u})$ for all $\mathbf{u}\in\{0,1\}^d$. With the multilinear extension, we can write the verification of arithmetic operations between vectors equivalently as the verification of sum of multi-variable, low-degree polynomial $g$:
\begin{equation}
    \label{equ: sum of poly}
    H\overset{\text{?}}{=}\sum_{(x_1,x_2,...x_v)\in \{0,1\}^v} g(x_1,x_2,...,x_v),
\end{equation}
by considering the multi-linear extension of tensors. 

\medskip
\textbf{Sum-Check/GKR Protocol.}
Algorithm \ref{alg:sumcheck protocol} describes the Sum-Check protocol, a.k.a. the GKR protocol~\cite{goldwasser2015delegating}, for Equation \eqref{equ: sum of poly}. The protocol proceeds in $v$ rounds~\cite{thaler2022proofs}. In the first round, the prover sends a polynomial $g_1(X_1)$ and claims it to be
\begin{equation}
\label{equ: proof of g_1}
    g_1(X_1)\overset{\text{?}}{=}\sum_{(x_2,...,x_v)\in\{0,1\}^{v-1}} g(X_1,x_2,...,x_v),
\end{equation}
where we use the capital letter (e.g., $X_1$) to represent the argument of the polynomial. A key observation is that if the polynomial $g_1(X_1)$ is as claimed in \eqref{equ: proof of g_1} and $H$ is as claimed in \eqref{equ: sum of poly}, then $H=g_1(0)+g_1(1)$. If so, the remaining proofs then proceed by proving Equation \eqref{equ: proof of g_1}. By the Schwartz-Zippel Lemma, it suffices for the prover to prove that Equation \eqref{equ: proof of g_1} holds at a random point $r_1\sim\mathbb{F}$:
\begin{equation}
\label{equ: proof of g_1(r_1)}
    g_1(r_1)\overset{\text{?}}{=}\sum_{(x_2,...,x_v)\in\{0,1\}^{v-1}} g(r_1,x_2,...,x_v).
\end{equation}
In the second round, the prover sends a polynomial $g_2(X_2)$ and claims it to be
\begin{equation}
\label{equ: proof of g_2}
    g_2(X_2)\overset{\text{?}}{=}\sum_{(x_3,...,x_v)\in\{0,1\}^{v-2}} g(r_1,X_2,x_3,...,x_v).
\end{equation}
Observe that if the polynomial $g_2(X_2)$ is as claimed in \eqref{equ: proof of g_2} and $g_1(r_1)$ is as claimed in \eqref{equ: proof of g_1(r_1)}, then $g_1(r_1)=g_2(0)+g_2(1)$.
By the Schwartz-Zippel Lemma, it suffices for the prover to prove that Equation \eqref{equ: proof of g_2} holds at a random point $r_2\sim\mathbb{F}$:
\begin{equation}
\label{equ: proof of g_2(r_2)}
    g_2(r_2)\overset{\text{?}}{=}\sum_{(x_3,...,x_v)\in\{0,1\}^{v-2}} g(r_1,r_2,x_3,...,x_v).
\end{equation}
In the third round, the prover sends a polynomial $g_3(X_3)$ and claims it to be
\begin{equation}
\label{equ: proof of g_3}
    g_3(X_3)\overset{\text{?}}{=}\sum_{(x_4,...,x_v)\in\{0,1\}^{v-3}} g(r_1,r_2,X_3,x_4,...,x_v).
\end{equation}
The recursive argument then continues with the proof of
\begin{equation}
\label{equ: proof of g_4}
    g_k(X_k)\overset{\text{?}}{=}\sum_{(x_{k+1},...,x_v)\in\{0,1\}^{v-k}} g(r_1,...,r_{k-1},X_k,x_{k+1},...,x_v).
\end{equation}
In the final round, all $\overset{\text{?}}{=}$'s can be proved w.h.p. (i.e. with failure probability at most $\frac{d}{|\mathbb{F}|}$, where $d$ is the degree of $g$) by the recursive argument if and only if $g_v(r_v)=g(r_1,r_2,...,r_v)$. The latter can be easily verified by the verifier via the opening of the commitment to $g$.

\begin{figure}[!h]
    \centering
    \includegraphics[width=0.85\textwidth]{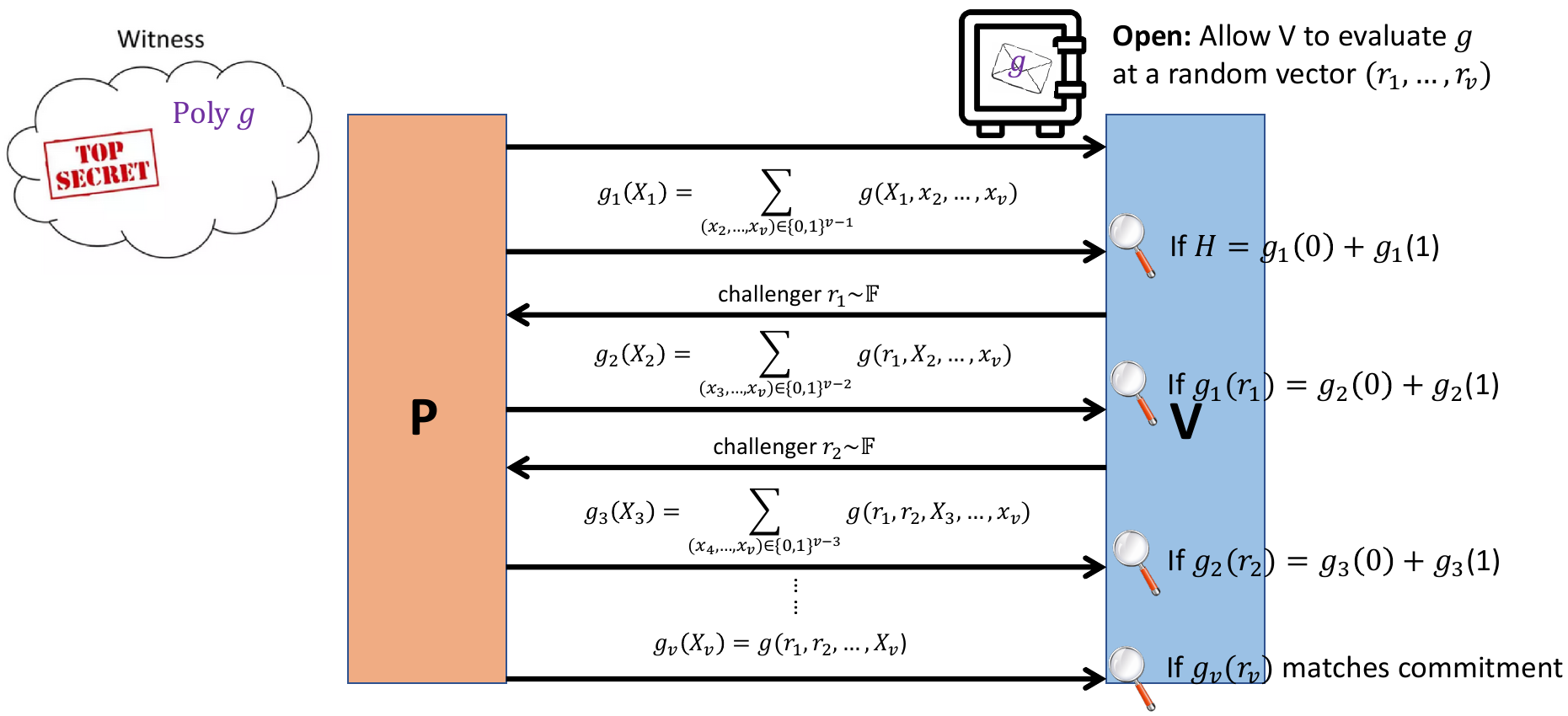}
    \caption{Sum-Check protocol for $H\overset{\text{?}}{=}\sum_{(x_1,x_2,...x_v)\in \{0,1\}^v} g(x_1,x_2,...,x_v)$.}
    \label{fig: sumcheck}
\end{figure}

The procedure of the Sum-Check protocol is shown in Figure \ref{fig: sumcheck}. The completeness is straightforward by the construction. The soundness follows from the recursive application of the Schwartz-Zippel Lemma for each round. The zero-knowledge property follows from the privacy guarantee of the Pedersen commitment.

\begin{algorithm}[!b]
\caption{Sum-Check protocol (interactive) for Equation \eqref{equ: sum of poly}}\label{alg:sumcheck protocol}
\begin{algorithmic}[1]
\State \textbf{Commitment:} Prover commits to the polynomial $g$ and sends the Pedersen commitment of $g$ to verifier (with an ability to prove the evaluation $g(r_1,r_2,...,r_v)$ at any point $(r_1,r_2,...,r_v)$, see Algorithm \ref{alg:commitment}).
\State Prover computes the polynomial with the argument $X_1$: $g_1(X_1)=\sum_{(x_2,...,x_v)\in\{0,1\}^{v-1}} g(X_1,x_2,...,x_v)$ and sends the polynomial to verifier.
\If{Verifier finds $H\not=g_1(0)+g_1(1)$}
    \State Verifier \textbf{outputs} \textsf{REJECT} and \textbf{ends} the algorithm.
\EndIf
\State Verifier samples a random number $r_1\in\mathbb{F}$ and sends it to prover.
\For{$i = 2,3,...,v$}
    \State Prover computes the polynomial with the argument $X_i$: $g_i(X_i)=\sum_{(x_{i+1},...,x_v)\in\{0,1\}^{v-i}} g(r_1,...,r_{i-1},X_i,x_{i+1},...,x_v)$ and sends the polynomial $g_i(X_i)$ to the verifier.
    \If{Verifier finds $g_{i-1}(r_{i-1})\not=g_i(0)+g_i(1)$}
        \State Verifier \textbf{outputs} \textsf{REJECT} and \textbf{ends} the algorithm.
    \EndIf
    \State Verifier samples a random number $r_i\in\mathbb{F}$ and sends it to prover.
\EndFor
\If{Verifier checks that $g_v(r_v)$ matches the opening of the commitment by a proof system in Algorithm \ref{alg:commitment}}
    \State Verifier \textbf{outputs} \textsf{PASS}.
\Else
    \State Verifier \textbf{outputs} \textsf{REJECT}.
\EndIf
\end{algorithmic}
\end{algorithm}

\medskip
\textbf{Reducing Linear Layers to Sum-Check Protocol.} Linear layers frequently appear in foundation models. Mathematically, linear layers can be written as matrix multiplication. Given $n\times n$ matrices $\mathbf{A}$ and $\mathbf{B}$ and we denote $\mathbf{A}\mathbf{B}$ as $\mathbf{C}$. One can interpret $\mathbf{A},\mathbf{B},\mathbf{C}$ as function $f_A,f_B,f_C:\{0,1\}^{\log_2 n}\times \{0,1\}^{\log_2 n}\rightarrow \mathbb{F}$:
\begin{equation}
    f_A(i_1,...,i_{\log_2 n},j_1,...,j_{\log_2 n})=A_{ij},
\end{equation}
where sequence $(i_1,...,i_{\log_2 n})$ and $(j_1,...,j_{\log_2 n})$ are the binary representations of $i$ and $j$, respectively.
Let $\widetilde{f_A},\widetilde{f_B},\widetilde{f_C}:\mathbb{F}^{\log_2 n}\times \mathbb{F}^{\log_2 n}\rightarrow \mathbb{F}$ denote the multilinear extension of $f_A,f_B,f_C$. It is easy to check that
\begin{small}
\begin{equation}
\label{equ: sumcheck for matrix product}
    \widetilde{f_C}(i_1,...,i_{\log_2 n},j_1,...,j_{\log_2 n})=\sum_{\mathbf{b}\in\{0,1\}^{\log_2 n}}\widetilde{f_A}(i_1,...,i_{\log_2 n},\mathbf{b})\cdot \widetilde{f_B}(\mathbf{b},j_1,...,j_{\log_2 n}).
\end{equation}
\end{small}
Equation \eqref{equ: sumcheck for matrix product} is in the form of \eqref{equ: sum of poly}. Thus, we can apply the Sum-Check protocol for verifying the execution of linear layers with $\log_2 n$-variant, degree-2 polynomial $g(\mathbf{b})=\widetilde{f_A}(i_1,...,i_{\log_2 n},\mathbf{b})\cdot \widetilde{f_B}(\mathbf{b},j_1,...,j_{\log_2 n})$ and $H=\widetilde{f_C}(i_1,...,i_{\log_2 n},j_1,...,j_{\log_2 n})$ at a random point $(i_1,...,i_{\log_2 n},j_1,...,j_{\log_2 n})\in\mathbb{F}^{\log_2 n\times \log_2 n}$.

\subsubsection{Proofs for Non-Arithmetic Operations}

There are two major techniques for zero-knowledge proof of non-arithmetic operations: 1) bit decomposition and 2) lookup table.

\medskip
\textbf{Reducing ReLU Activation to Bit Decomposition.} Bit decomposition is one of the most frequently used techniques for zero-knowledge proofs of non-arithmetic operations. Let $\mathbf{Z}$ denote the pre-activated tensor of ReLU, and let $\mathbf{A}$ denote the after-activated tensor. We now introduce the techniques appearing in \cite{sun2023zkdl}. In the ReLU activation, we have
\begin{equation}
    \label{equ: ReLU}
    \mathbf{A} = \text{sign}(\mathbf{Z}) \odot \text{abs}(\mathbf{Z}),
\end{equation}
where $\odot$ is the Hadamard product, $\text{abs}(z)$ is the element-wise absolute value of $\mathbf{Z}$, and
\begin{equation}
    \text{sign}(z)=
    \begin{cases}
        0, & \text{if $z<0$};\\
        1, & \text{otherwise},\\
    \end{cases}
\end{equation}
is applied element-wise to the tensor $\mathbf{Z}$. Denote by $z=(z_0,z_1,...,z_Q)$ the Q-bit binary representation of $z$, where $Q=\log_2 p$ (e.g., 128 or 256), $p$ is the (prime) order of the finite field $\mathbb{F}$, $z_0=\text{sign}(z)\in\{0,1\}$ represents the sign, and $(z_1,...,z_Q)=\text{abs}(z)$ refers to the Q-bit unsigned binary representation of $z$. To prove Equation \eqref{equ: ReLU} holds for all elements $z$, we only need to prove: 1) $(z_0,z_1,...,z_Q)$ is binary representation such that all elements are in $\{0,1\}$: $0\overset{\text{?}}{=}z_i(z_i-1)$ for $i\in\{0,1,...,Q\}$; 2) $(z_0,z_1,...,z_Q)$ can recover $z$ in the sense that
\begin{equation}
    z\overset{\text{?}}{=}\underbrace{(2\times z_0-1)}_{\text{sign}}\times \underbrace{(2^{Q-1}\times z_1+2^{Q-2}\times z_2+...+2^0\times z_Q)}_{\text{Q-bit unsigned integer}};
\end{equation}
and 3) Equation \eqref{equ: ReLU} is correctly executed for element $a\in\mathbf{A}$ in the sense that
\begin{equation}
a\overset{\text{?}}{=} z_0\times (2^{Q-1}\times z_1+2^{Q-2}\times z_2+...+2^0\times z_Q).
\end{equation}
Therefore, with bit decomposition, we can reduce the non-arithmetic ReLU operation to the arithmetic operations in the proofs of 1) and 2) and call the Sum-Check protocol. We can also use the Schwartz-Zippel Lemma to merge all the proofs in 1) and 2) into a single one:
\begin{equation}
    \label{equ: proof compression}
    \begin{split}
    &(2z_0-1)(2^{Q-1}z_1+...+2^0 z_Q)-z+(z_0(2^{Q-1}z_1+...+2^0 z_Q)-a)r\\
    &+\sum_{i=0}^Q z_i(z_i-1)r^{i+2} \overset{\text{?}}{=} 0,
    \end{split}
\end{equation}
where $r\sim\mathbb{F}$ is a random sample from the finite field $\mathbb{F}$.

\medskip
\textbf{Reducing Non-Arithmetic Operations to Lookup Tables.} A lookup table is another technique for zero-knowledge proofs of non-arithmetic operations. It resolves the issue in the bit decomposition approach that the binary representation is lengthy when $p$ is large, so one non-arithmetic operation is represented by as many as $\log_2 p$ bit-operations~\cite{thaler2022proofs}. Hereby, the length of $\log_2 p$ comes from the use of base-2 in the bit decomposition. In order to resolve the issue, one could consider a larger base $k$ and check the correction of base-$k$ decomposition by a table lookup. We now introduce the techniques appearing in \cite{sun2024zkllm}. In particular, both the prover and the verifier parties maintain a shared table: $\mathcal{T}:=\{(\mathbf{T}_\mathcal{X},f(\mathbf{T}_\mathcal{X}))\}\in\mathbb{F}^{n_2\times (d_1+d_2)}$ for a non-arithmetic operation $f$ (e.g., the Softmax operation). To reduce the non-arithmetic operation $f$ to a lookup table argument, the prover needs to convince the verifier that the $d_1$-dim input vector $\mathbf{X}$ and the $d_2$-dim output vector $\mathbf{Y}$ satisfy $\sum_{i=0}^{d_1-1} r^i\mathbf{X}_i+\sum_{i=0}^{d_2-1} r^{i+d_1}\mathbf{Y}_i \subset \sum_i \textsf{Col}_i(\mathcal{T}) r^i$ for a random $r\sim\mathbb{F}$, where $\textsf{Col}_i(\mathcal{T})$ refers to the $i$-th column of table $\mathcal{T}$. One can further reduce the proof of this ``subset'' relations to a Sum-Check proof, by observing that: $\mathbf{S}\subset \mathbf{T}$ for $\mathbf{S}\in \mathbb{F}^{n_1}$ and $\mathbf{T}\in \mathbb{F}^{n_2}$, if and only if there exists $\mathbf{e}=(e_0,e_1,...,e_{n_2-1})\in\mathbb{F}^{n_2}$ such that the following two polynomials over $X$ is identical:
\begin{equation}
    \prod_{i\in[n_1]} (X+\mathbf{S}_i) = \prod_{i\in[n_2]} (X+\mathbf{T}_i)^{e_i}.
\end{equation}
By taking the logarithmic derivative at both sides, the following two rational functions are identical:
\begin{equation}
\label{equ: rational function}
    \sum_{i\in[n_1]} \frac{1}{X+\mathbf{S}_i} = \sum_{i\in[n_2]} \frac{e_i}{X+\mathbf{T}_i}.
\end{equation}
The prover sets and commits to $e_i=|\{j: \mathbf{S}_j=\mathbf{T}_i\}|$.
Equation \eqref{equ: rational function} can be proved by the Sum-Check protocol with a random variable $X\sim \mathbb{F}$, as all the operations here are arithmetic~\cite{sun2024zkllm}.

\subsubsection{Zero-Knowledge Proofs for Foundation Models}

\begin{table}[h!]
\centering
\caption{Zero-knowledge proof techniques for various operations in foundation models.}
\label{table: zkp techniques}
\scalebox{0.97}{
\begin{tabular}{ c|c }
 \hline
 Operations & Zero-Knowledge Proof Techniques \\
 \hline
 Linear Layer~\cite{thaler2022proofs} & Sum-Check Protocol \\ 
 Convolution~\cite{liu2021zkcnn} & Sum-Check Protocol/FFT \\ 
 Residual Connection~\cite{sun2024zkllm} & Sum-Check Protocol \\ 
 Softmax/Sigmoid/SwiGLU/GELU/GLU~\cite{sun2024zkllm} & Lookup Table \\ 
 LayerNorm/RMSNorm~\cite{sun2024zkllm} & Lookup Table \\
 ReLU~\cite{sun2023zkdl} & Bit Decomposition \\
 \hline
\end{tabular}}
\end{table}

By integrating all components, one can construct a zero-knowledge proof system for foundation models like Meta's LLaMA series. Foundation models comprise transformer layers followed by MLP layers. For arithmetic operations, such as the linear layers in transformers and MLPs, the Sum-Check protocol can be effectively employed to generate proofs. For non-arithmetic operations, like Softmax and SwiGLU activation, a lookup table can be utilized. For piece-wise linear activation, such as ReLU, bit decomposition can be used. Table \ref{table: zkp techniques} lists the zero-knowledge proof techniques for different operations in foundation models. Note that each operation requires carefully designed techniques for efficiency. To the best of our knowledge, zkLLM~\cite{sun2024zkllm} is the first work that introduces zero-knowledge proofs to foundation models with tens of billions of parameters such as OPT~\cite{zhang2022opt} and LLaMA-2~\cite{hugo2023llama2}.

\subsubsection{zkDPS: Zero-Knowledge Decentralized Proof System}

We build a zero-knowledge system, zkDPS, for decentralized large language models. zkDPS is built upon zkLLM~\cite{sun2024zkllm} and zkDL~\cite{sun2023zkdl}. zkLLM~\cite{sun2024zkllm} and zkDL~\cite{sun2023zkdl} are state-of-the-art CUDA implementation of zero-knowledge proofs of machine learning. However, speed remains the most significant pain point for zero-knowledge proofs of foundation models. For example, it takes zkLLM roughly 15 minutes to generate a proof for the inference procedure of LLaMA-2 13B with an input token length of 2,048~\cite{sun2024zkllm}. In comparison, vanilla inference of zkLLM takes less than 0.1 seconds per token. Despite significant efforts have been made to accelerate generic zero-knowledge proof frameworks such as zk-STARK and zk-SNARK~\cite{chen2022review}, they are still not fast enough, even much slower than zkLLM for billion-scale models. Therefore, the acceleration of zero-knowledge proofs is urgent for real-world applications.

zkDPS proposes the following ideas to speed up proof generation beyond zkLLM~\cite{sun2024zkllm} and zkDL~\cite{sun2023zkdl}. One idea could be to leverage the ``compressible property'' of LLMs by algorithmic innovation. It is well-known that LLMs are in essence a compression of information and data from the real world. Thus one could consider using a small draft model to ``guess'' the output of the original LLM, similar to speculative sampling. As the draft model is small (typically 20x-50x smaller than the original LLM~\cite{li2024eagle}), one could expect to experience a shorter time for proof generation of the execution of the draft model. Experimentally, the top-1 guessing accuracy of the draft model is as high as 82\% in EAGLE~\cite{li2024eagle,li2024eagle2}, a state-of-the-art speculative sampling method. Therefore, this acceleration technique will not degrade the inference performance of LLMs too much.

Another idea is to leverage the hardware acceleration of modern GPUs. zkLLMs has already used CUDA implementation to parallelize the computations in the proof generation. For example, the CUDA kernels such as Pedersen commitment, elliptic curve, and lookup tables have been re-implemented or adapted in zkLLMs~\cite{sun2024zkllm,sun2023zkdl}, making it the fastest implementation of zero-knowledge large language models at the submission of this paper. We are optimizing those kernels by engineering methods such as Tensor Parallelism and KV cache.

\section{Security and Privacy for General Inference}
\label{sec:general-inference-appx}

In general inference scenarios, the results of AI inference tasks are less critical, allowing for faster processing speeds and lower costs without compromising basic security and privacy standards. These tasks include everyday activities such as checking the weather, recommending products, or filtering spam emails. Our strategy for ensuring model verification and integrity in these scenarios is Consensus-based Verification Check (CBV), which leverages the collective agreement of multiple nodes to ensure the correctness and integrity of model execution without revealing sensitive data. This approach is detailed in Section \ref{subsec:cbv}. 
For data privacy, we employ split learning (SL) \cite{vepakomma2018split,thapa2022splitfed}, a method where the model is divided into segments, and each segment is trained on different nodes to maintain data privacy by ensuring no single node has access to the complete dataset. More information on this technique can be found in Section \ref{subsec:sl}.
Fig. \ref{fig:general-flow} provides an overview of the security flow of general inference, which first applies SL to encrypt the first (few) layers to ``encrypt'' the raw data, where the AI models are sharded and deployed to nodes with redundancy. For the nodes who have the same shard, their outputs will be verified by consensus for integrity.

\subsection{Consensus-based Verification Check for Model Integrity}
\label{subsec:cbv}

Given the computational and scalability challenges associated with ZKML for verifying the integrity of LLMs in decentralized systems, 
We proposes a consensus-based distribution verification (CDV) strategy for general inference scenarios. 
This strategy leverages the collective agreement of multiple nodes to ensure the correctness and integrity of model execution without revealing sensitive data.

\textbf{Consensus-based Verification.} 
Consider a decentralized network with $N$ nodes, where each node $i$ executes the same inference model $\mathcal{M}$ with parameters $\theta$, on a given input $x$  \cite{chen2021robust}. The output of the model on node $i$ is denoted by $y_i = \mathcal{M}(x; \theta_i)$. The goal is to ensure that all nodes accurately execute the model $\mathcal{M}$, yielding consistent outputs.
The process can be formalized in the following steps:

\begin{enumerate}
    \item \textbf{Redundant execution.} A subset of the network nodes, $\{1, 2, \ldots, k\} \subseteq N$, independently computes the output $y_i$ for the same input $x$.
    \begin{equation}
        y_i = \mathcal{M}(x; \theta), \quad \forall i \in \{1, 2, \ldots, k\}.
    \end{equation}
    
    \item \textbf{Output collection.} The outputs $\{y_1, y_2, \ldots, y_k\}$ are collected for consensus evaluation. This collection phase requires secure and efficient communication protocols to protect the integrity of the transmitted data.
    
    \item \textbf{Consensus determination.} Utilizing a consensus algorithm $\mathcal{C}$, the system evaluates the collected outputs to determine the agreed-upon result $y_{\text{con}}$. The consensus result is considered valid if it satisfies a predefined criterion, such as majority agreement or a more sophisticated decision rule based on the specific properties of the outputs.
     \begin{equation}
     y_{\text{con}} = \mathcal{C}(\{y_1, y_2, \ldots, y_k\}).
     \end{equation}
    
    \item \textbf{Verification and finalization.} If the consensus results $y_{\text{con}}$ align with the outputs from a sufficiently large subset of nodes, the model's execution is verified. Otherwise, discrepancies indicate potential integrity issues, triggering further investigation or corrective measures.
\end{enumerate}

This consensus-based approach not only facilitates the verification of model integrity across decentralized nodes but also introduces a robust mechanism to detect and mitigate the impact of faulty or malicious nodes.

\textbf{Taking Model Sharding into Account.} In our decentralized system, where ML models' computational graphs may be sharded across multiple nodes for scalability, each node $i$ possesses a unique shard $\mathcal{M}_i$ of the complete model $\mathcal{M}$. This partitioning requires a specialized approach to Consensus-based Verification to accommodate the fragmented nature of model execution.

Consider the complete model $\mathcal{M}$ being divided into $k$ shards, such that $\mathcal{M} = \bigoplus_{i=1}^{k} \mathcal{M}_i$, where $\bigoplus$ denotes the operation of combining the model shards to represent the full model functionality. Given an input $x$, the execution of these shards across $k$ nodes produces a set of partial outputs $\{y_1, y_2, \ldots, y_k\}$, where $y_i = \mathcal{M}_i(x; \theta_i)$.

\textbf{Verification in the Context of Sharding.}
\begin{enumerate}
    \item \textbf{Shard Redundant Execution.} For each shard $\mathcal{M}_i$ of the complete model $\mathcal{M}$, redundant execution is performed by a designated subset of nodes. Each of these nodes, within the subset responsible for shard $\mathcal{M}_i$, computes the output $y_{i,j}$ for the given input $x$, where $j$ represents the node within the subset.
    \begin{equation}
    y_{i,j} = \mathcal{M}_i(x; \theta_{i,j}), \quad \forall j \in \text{Subset of nodes for } \mathcal{M}_i.
    \end{equation}
    This step introduces computational redundancy, where multiple independent computations of the same shard aim to fortify the verification process by cross-verifying results among nodes computing the same shard.
    
    \item \textbf{Redundant Output Collection and Verification.} The outputs $\{y_{i,1}, y_{i,2}, \ldots, y_{i,m}\}$ for each shard $i$ are collected from the nodes in its subset. A consensus mechanism $\mathcal{C}_i$ specific to shard $i$ then evaluates these collected outputs to determine a shard-specific agreed-upon result $y_{\text{con},i}$.
    \begin{equation}
    y_{\text{con},i} = \mathcal{C}_i(\{y_{i,1}, y_{i,2}, \ldots, y_{i,m}\}).
    \end{equation}
    Here, $m$ denotes the number of nodes executing the shard $\mathcal{M}_i$. The redundancy in computation across these nodes allows for a robust verification mechanism, enhancing the detection of discrepancies or faults.
    
    \item \textbf{Shard Verification Completion.} Upon achieving consensus for a shard $i$, signified by the result $y_{\text{con},i}$, the process ensures the integrity of the shard's computation before proceeding. This step-by-step verification across shards, with redundancy in each shard's computation, significantly reduces the risk of erroneous or malicious model execution.
    
    \item \textbf{Model Reconstruction.} After each shard has been independently verified, the shard-specific consensus results \\
    $\{y_{\text{con},1}, y_{\text{con},2}, \ldots, y_{\text{con},k}\}$ are combined to reconstruct the final model output $Y_{\text{final}}$. This comprehensive output can ensure the integrity of the complete model execution.
    \begin{equation}
    Y_{\text{final}} = \bigoplus_{i=1}^{k} y_{\text{con},i}.
    \end{equation}
\end{enumerate}

\textbf{Consensus-based Distribution Verification (CDV).} Building upon traditional consensus mechanisms, the CDV strategy introduces an advanced layer of verification by assessing the statistical distribution of model outputs across a decentralized network. 
This approach is ideally suited for scenarios where the model is not monolithic but is instead distributed as shards across multiple nodes.

CDV is based on the understanding that while individual outputs from model shards might exhibit slight variability due to the stochastic nature of ML models and the complexity of input data, the collective output distribution should maintain consistency. This consistency holds, provided that the model and its inputs remain unchanged. 
By evaluating the aggregated statistical characteristics of these outputs, CDV furnishes a sophisticated and robust framework for affirming the uniformity and integrity of the model's behavior, thereby enhancing security and privacy without direct comparison of individual inference results.

\begin{itemize}
    \item \textbf{Sharded Execution and Output Synthesis.} In the initial phase, each node, housing a shard $\mathcal{M}_i$ of the overarching model $\mathcal{M}$, executes its segment on a shared input $x$, generating partial outputs $\{y_1, y_2, \ldots, y_k\}$. These outputs are synthesized to construct a comprehensive output profile reflecting the entire model's combined inference result.
    
    \item \textbf{Advanced Statistical Aggregation.} Following output synthesis, the system embarks on advanced statistical analysis, deriving metrics such as the mean $\mu$, standard deviation $\sigma$, and potentially higher-order moments. This stage may also incorporate non-parametric statistics to capture the full essence of the output distribution, offering a nuanced view of the model's performance landscape.
    
    \item \textbf{Rigorous Distribution Comparison.} Utilizing sophisticated statistical methodologies, the derived metrics are juxtaposed with predefined benchmarks or dynamically established norms. Techniques such as hypothesis testing, divergence measures, or similarity indices evaluate the congruence between the observed and expected output distributions, facilitating an objective assessment of model integrity.
    
    \item \textbf{Enhanced Consensus Mechanism with Adaptive Thresholding.} The core of CDV lies in its consensus mechanism, where nodes collectively determine the acceptability of the observed distribution's alignment with benchmarks. Adaptive thresholding plays a crucial role here, dynamically adjusting sensitivity based on historical data and operational context to pinpoint deviations that truly signify integrity breaches.
\end{itemize}

Through its implementation, CDV offers a powerful solution to the challenges of verifying the integrity of distributed ML models in our decentralized framework. 
By focusing on distributional characteristics rather than discrete output values, CDV not only elevates the verification process but also aligns with the goals of enhancing model security and maintaining stringent privacy standards.

\subsection{Data Privacy Protection via Split Learning}
\label{subsec:sl}
Recognizing the challenges posed by encrypting data for use in decentralized inference systems, we adopts Split Learning (SL) as a pragmatic solution to facilitate secure and efficient computation on encrypted data \cite{pham2023binarizing,khan2023split}.
Traditional encryption methods such as HE, while securing data at rest and in transit, render it costly for direct computation by obscuring its format and structure. 
This limitation is particularly problematic for processing with LLMs within a decentralized framework, where data privacy cannot be compromised.

Split Learning \cite{vepakomma2018split} addresses these concerns by partitioning the computational model, allowing for data to be processed in parts without revealing sensitive information. 
In essence, the user data is protected by not being directly transmitted to any nodes -- only the data embeddings are being passed around, and each node will only be accessing the embeddings of certain layers.

Consider a neural network model \( \mathcal{N} \), such as Llama 2 \cite{touvron2023llama} composed of a sequence of 32 layers \( \{L_1, L_2, \ldots, L_{32}\} \), each with its own set of parameters \( \Theta_i \) and activation function \( \sigma_i \). 
The input to the network is \( X \), and the output of the \(i\)-th layer, given input \( x_i \), can be mathematically described as:

\begin{equation}
a_i = L_i(x_i; \Theta_i) = \sigma_i(W_i x_i + b_i),
\end{equation}

where \( W_i \) and \( b_i \) are the weight matrix and bias vector of the \(i\)-th layer, respectively, and \( \sigma_i \) is a nonlinear activation function such as ReLU, sigmoid, or tanh.

Assuming the model is split at layer \( k \), where the client handles layers \( \{L_1, \ldots, L_k\} \) and the server handles layers \( \{L_{k+1}, \ldots, L_{32}\} \). The client computes the intermediate representation \( Z \) as follows:

\begin{equation}
Z = \sigma_k(W_k \cdot \sigma_{k-1}( \ldots \sigma_1(W_1 X + b_1) \ldots ) + b_k).
\end{equation}

This intermediate representation \( Z \) is then transmitted to the server, which continues the computation:

\begin{equation}
Y = \sigma_{32}(W_{32} \cdot \sigma_{31}( \ldots \sigma_{k+1}(W_{k+1} Z + b_{k+1}) \ldots ) + b_{32}).
\end{equation}

The loss function \( \mathcal{L}(Y, Y_{true}) \) computes the error between the network output \( Y \) and the true labels \( Y_{true} \), and the gradient of the loss with respect to the model's parameters through backpropagation:

\begin{equation}
\frac{\partial \mathcal{L}}{\partial \Theta_i} = \text{ChainRule}\left(\frac{\partial \mathcal{L}}{\partial Y}, \frac{\partial Y}{\partial a_{32}}, \ldots, \frac{\partial a_i}{\partial \Theta_i}\right).
\end{equation}

For privacy concerns during the transmission of \( Z \) from client to server, differential privacy methods may be applied \cite{dwork2006differential}. Defining a privacy metric \( \mathcal{P} \) that quantifies the information leakage from the intermediate representation \( Z \), a proof of privacy preservation could demonstrate that for any \( \epsilon \)-differential privacy guarantee, the information leakage remains below a threshold:

\begin{equation}
\mathcal{P}(Z) \leq \epsilon.
\end{equation}

It is noted that by using differential privacy with SL, the privacy will be improved at the cost of inference quality \cite{behnia2022ew}. Thus, in our framework, this is defined as a tunable parameter to be decided, given the user requirements.

By leveraging Split Learning, we effectively navigates the complexities of data encryption within its decentralized inference system for LLMs. 
This approach not only preserves the confidentiality and integrity of user data but also ensures the operational feasibility of complex model computations, demonstrating a sophisticated balance between privacy preservation and computational pragmatism.

\section{Our Specialized Trusted Execution Environments for Security and Privacy}
\label{sec:tee}

Trusted Execution Environments (TEEs) effectively establish security and privacy within our decentralized inference architecture. 
Using TEEs, we can create secure isolation zones within the network's nodes, guaranteeing the privacy and accuracy of data and computational operations. In our scenarios, two key pieces of information should be protected from the node runners: (i) users' input data for AI inference and (ii) private model parameters, where TEE provides a hardware-based solution for it.

To ensure the protection of sensitive processes and data within nodes from unauthorized access, including the node operators themselves, TEEs provide an aggressive isolation mechanism. The significance of isolation is of utmost importance in a decentralized setting where the inference of AI models is spread among multiple owners.

\begin{enumerate}
\item \textbf{Isolated Execution}. TEEs provide a secure enclave for executing computations, isolated from the rest of the node’s operating environment. This isolation ensures that even if other parts of the node are compromised, the computations within the TEE remain protected. When nodes process only fragments of a model, ensuring the integrity and confidentiality of each fragment is crucial. TEEs prevent any unauthorized access or tampering with the code and data inside the enclave, thus safeguarding the partial model computations.

\item \textbf{Data Privacy}. Since each node handles only parts of the AI model, sensitive data processed by the model can potentially be less secure if exposed to less protected parts of the system or network. TEEs encrypt the data within the enclave, ensuring that any information processed remains confidential. Even if data needs to be shared across nodes, it can be encrypted before leaving the TEE, ensuring that it travels securely across the network.

\item \textbf{Consistency and Integrity}. TEEs can ensure the integrity of the computations performed on each node. Through cryptographic techniques such as attestation, a TEE can prove to other nodes or external verifiers that the computations were performed correctly without revealing the underlying data. This attestation process allows other nodes in the decentralized network to trust the results from each node without having to access the actual data or model details, thus maintaining consistency and integrity across the decentralized model.

\end{enumerate}

\textbf{CPU- and GPU-based TEEs.}
Intel, AMD, and NVIDIA all provide TEEs, while NVIDIA uniquely supports GPU TEEs. 
This capability is crucial for accelerating AI and high-performance computing tasks securely, where GPUs can efficiently handle tensor-based operations.
The general procedure involves initializing a CPU TEE virtual machine (VM) first, which then controls the GPU TEE (available only with NVIDIA\footnote{NVIDIA's Hopper architecture GPUs, such as the H100 Tensor Core GPUs, enable this secure execution environment, supporting a wide range of AI and high-performance computing use cases.
}).
This setup ensures that the owner of the CPU and GPU cannot see the contents of the VM, providing an additional layer of security. 
In other words, even though the node runner owns the hardware, they cannot access the AI model and data inside the VM. 


By leveraging TEEs, we can provide robust security and privacy measures that are essential for both critical and general inference scenarios, ensuring that sensitive data and model parameters are always protected from unauthorized access and tampering.
Thus, we design a specialized version of TEEs that best suits our scenarios with decentralized inference.
Note that our specialized TEE is feasible for all node runners with or without GPUs.

\textbf{The Advantages of TEEs} include:
\begin{itemize}
\item \textbf{Low Overhead and High Performance}: TEEs provide a secure computing environment with minimal performance overhead (i.e., can be as small as 3\% as reported in different literature \cite{volos2018graviton,mo2020darknetz}), ensuring efficient execution of AI inference tasks. This allows us to maintain high performance in both critical and general inference scenarios.

\item \textbf{Support for Both CPU and GPU}: TEEs can be implemented on both CPUs and GPUs, with the general procedure involving initializing a CPU TEE virtual machine (VM) that controls the GPU TEE (available only with NVIDIA). This setup, supported by NVIDIA H100 Tensor Core GPUs, ensures seamless interaction with the hardware and enhances the security of AI computations.

\item \textbf{Data and Model Security}: The hardware owner (node runner) will not have access to the AI model and data inside the VM. This ensures the confidentiality and integrity of the computations, protecting sensitive data and model parameters from unauthorized access, even by the hardware owners.

\item \textbf{Elimination of Algorithm-Based Approaches}: Utilizing TEEs negates the need for algorithm-based security measures like ZKML and HE discussed above in \S \ref{subsec:zkp}. This results in faster performance due to the inherent efficiency of hardware-based security, making it ideal for our decentralized AI inference tasks.

\item \textbf{Scalable and Secure Collaboration}: 
TEEs enable secure collaboration across multiple nodes by creating isolated enclaves that protect data during computation. This is particularly beneficial for our decentralized AI inference, allowing multiple organizations to pool their data securely for training and inference while maintaining data privacy and compliance.
\end{itemize}

\begin{figure}[!ht]
\centering
\begin{subfigure}[b]{0.48\textwidth}
   \includegraphics[width=\linewidth]{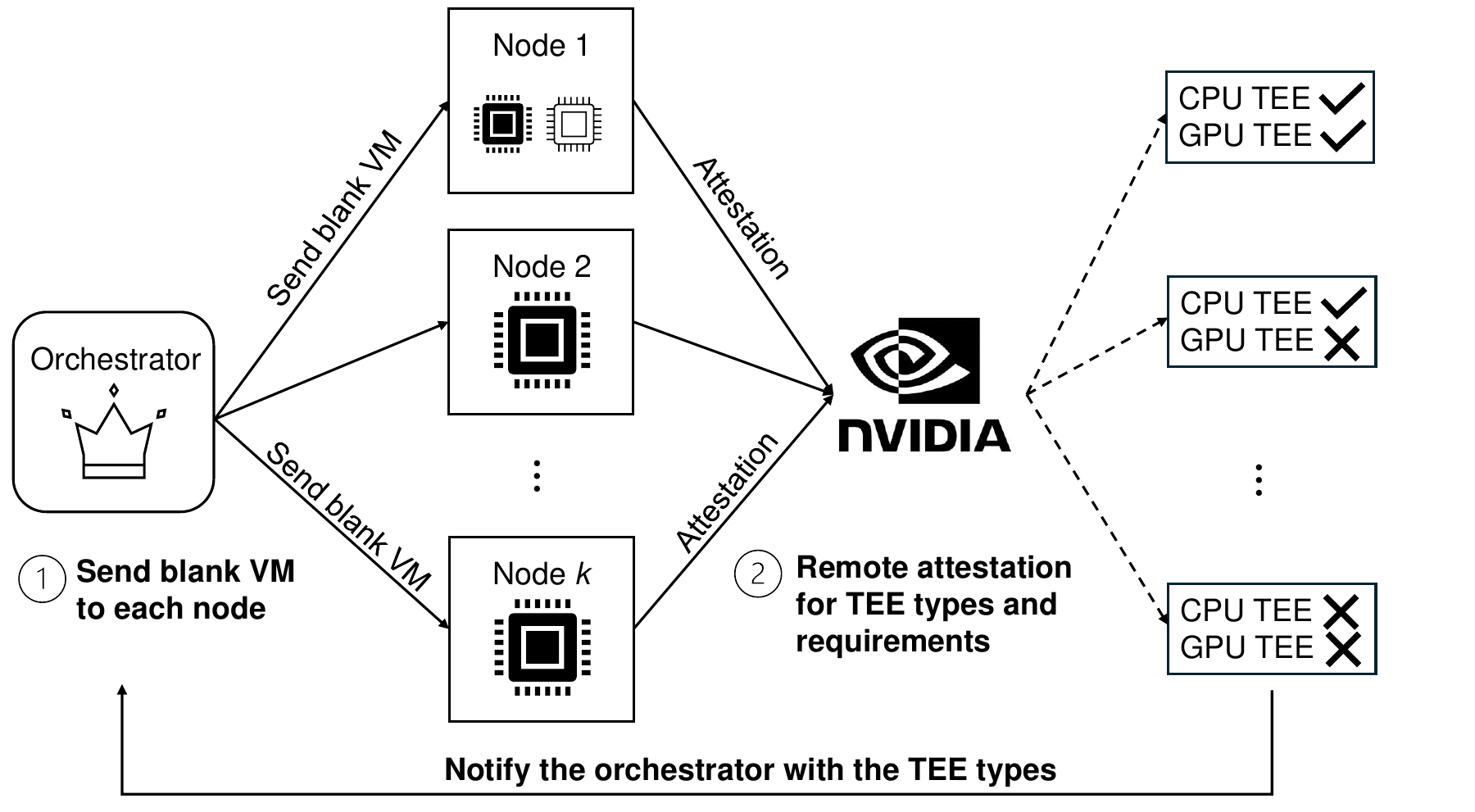}
   \caption{The first two steps are (1) sending an empty secure VM (without any data and models) to each node runner by the orchestrator and (2) running remote attestation with the server provider like NVIDIA to ensure the types of TEEs a node can support.}
   \label{fig:sub1}
\end{subfigure}
\hfill 
\begin{subfigure}[b]{0.48\textwidth}
   \includegraphics[width=\linewidth]{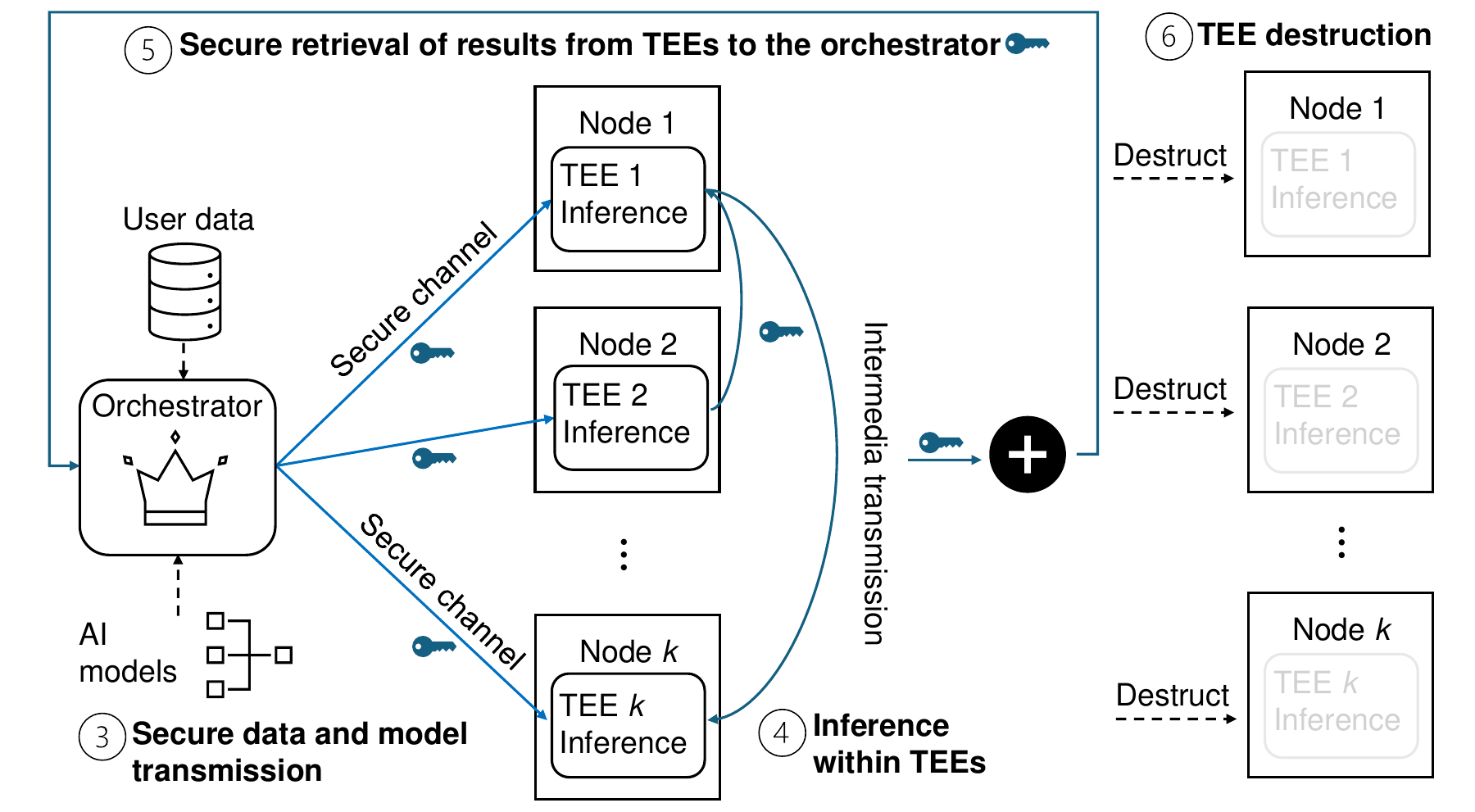}
   \caption{After knowing the TEE types each node can support, the orchestrator (3) sends user data and AI models to each node's TEEs via the \textcolor{cobalt}{secure channel}, after which (4) nodes execute AI inferences within TEEs and directly communicate with each other via \textcolor{cobalt}{secure channels}. Finally, The orchestrator (5) retrieves the aggregated results from TEEs securely and notifies the node runners to destroy each TEE.}
   \label{fig:sub2}
\end{subfigure}
\caption{Overview of our specialized TEE paradigm for decentralized AI inference.}
\label{fig:tee-overview}
\end{figure}

\subsection{Implementation of Our Specialized TEE in Detail}

As a decentralized AI provider, we aim to have multiple node runners collaboratively leverage their hardware to perform AI inference for users. In this setup, we employ one orchestrator within the Rich Execution Environment (REE) to facilitate the entire process within our specialized TEE. 
The REE, unlike the TEE, is the standard, general-purpose operating environment that handles regular applications and processes without the enhanced security measures of a TEE. 
We have the user data \(\mathbf{X}\) and the AI model \(M\) (e.g., Llama 2 \cite{touvron2023llama}), and we have \(k\) node runners for hosting TEEs.
Fig. \ref{fig:tee-overview} provides a high-level overview of the implementations of our specialized TEE.

\textbf{Step 1: TEE Initialization.} 
We first send a blank Virtual Machine (VM) TEE to each node runner. This VM does not contain any data or model information initially, ensuring no sensitive information is at risk during the setup phase.

\textbf{Step 2: Remote Attestation}.
The orchestrator initiates "remote attestation" with NVIDIA to verify that the node runner's machine meets the requirements for NVIDIA's TEE. Remote attestation involves the following steps:
\begin{enumerate}
    \item The attester (TEE) collects a set of claims that represent the state of its system, including a cryptographic hash of the application’s code:
    \[
    \text{Measurement} = H(\text{Code}),
    \]
    where \(H\) denotes a cryptographic hash function.
    \item These claims, along with the hardware state, are signed to form evidence:
    \[
    \text{Evidence} = \text{Sign}_{\text{Private Key}}(\text{Claims}).
    \]
    \item The evidence is sent to the verifier, who checks its validity using the public key:
    \[
    \text{Verify}(\text{Evidence}, \text{Public Key}) \rightarrow \text{True/False}.
    \]
    \item Based on the attestation results, the orchestrator determines whether each node supports GPU TEE or CPU TEE only:
    \[
    \text{TEE Type} = \begin{cases} 
    \text{GPU TEE}, & \text{if GPU TEE requirements met}; \\
    \text{CPU TEE}, & \text{if only CPU TEE requirements met}.
    \end{cases}
    \]
    If the machine meets the requirements for a GPU TEE, the NVIDIA toolkit is used to initialize the GPU VM. Otherwise, only a CPU-based VM is initialized.
\end{enumerate}

If the machine meets the requirements for a GPU TEE, the NVIDIA toolkit can be used to initialize the GPU VM. Otherwise, only a CPU-based VM is initialized.
The supported types of TEE information are sent back to the orchestrator.

\textbf{Step 3: Secure Data and Model Transmission}.
Once attestation confirms the TEE's security, we securely transmit the user data \(\mathbf{X}\) and the AI model fragment \(M_i\) to the node runner's TEE via an encrypted channel, similar to SSH:

\[
\text{Secure Transmission:} \quad \text{Enc}(\{\mathbf{X}, M_i\}) \xrightarrow{\text{SSH}} \text{TEE}.
\]

Here, \(\text{Enc}\) denotes encryption, ensuring that the data and model remain confidential and are not exposed to unauthorized access during transit.

\textbf{Step 4: AI Inference within TEEs.}
Each node runner's TEE performs a part of the AI inference using the model fragment \(M_i\) on the data \(\mathbf{X}\). The TEE provides an isolated environment for secure computation, ensuring the node runner cannot access the sensitive data or model parameters. The computation within the TEE can be denoted as:

\[
\text{Inference:} \quad Y_i = M_i(\mathbf{X}).
\]

For sequential tasks, the orchestrator sets up encrypted channels among the nodes with computation dependencies. Each node can directly send and receive intermediate inference results to and from other dependent nodes (the details are provided in the next section):

\[
\text{Intermediate Transmission:} \quad \text{Enc}(Y_i) \xrightarrow{\text{Encrypted Channel}} \text{Next Node}
\]

\textbf{Step 5: Secure Retrieval of Final Results.}
After the final computation is completed, the final inference results \(Y_k\) are securely transmitted back to the orchestrator from the last node's TEE using an encrypted channel like SSH:

\[
\text{Secure Retrieval:} \quad \text{Enc}(Y_k) \xrightarrow{\text{SSH}} \text{Orchestrator}
\]

\textbf{Step 6: Destruction of TEE.}
Once the inference results have been successfully retrieved and verified, the TEE on the node runner is destroyed. This step ensures that any residual data or model information is completely removed from the node runner's hardware. The process can be formalized as:

\[
\begin{aligned}
    &\text{Destroy TEE:} \\
    &\text{Step 1: Securely erase all data and model information}; \\
    &\text{Step 2: Shut down the TEE}.
\end{aligned}
\]

By implementing TEEs in this manner, we ensure that the AI inference process is secure, protecting both the user's data and the AI model throughout the entire lifecycle of the computation. This approach leverages the robustness of hardware-based security measures, providing high performance and strong guarantees of confidentiality and integrity. TEEs also facilitate mutual attestation, where both the user and the node runner can verify each other's integrity, further enhancing the trustworthiness of the decentralized AI system.

\subsection{Our Innovations in Specialized TEE}

Our system operates in a unique decentralized AI setting, distinct from the general usage of TEEs where only one REE interacts with one TEE to run a model or a specific task. 
Our system involves multiple TEEs across various nodes to run a shard of an AI model, requiring advanced strategies to ensure efficiency and security.

\textbf{Innovation 1: Communication Optimization Among Multiple TEEs}.
We optimize communication by minimizing data transfer between the orchestrator (considered as a REE) and each TEE. 
Instead of routing all communications through the orchestrator, we establish pre-set secure communication channels among the TEEs. 
This allows TEEs to communicate directly with each other, significantly reducing latency and improving overall system efficiency. 
Direct communication between TEEs is secured using encrypted channels to ensure data privacy and integrity during transit, and the owners of the nodes cannot see the communication either. 
This approach reduces bottlenecks and enhances the speed of decentralized inference processes, which is crucial for real-time applications.

\[
\text{Direct Communication:} \quad \text{TEE}_i \xrightarrow{\text{Encrypted Channel}} \text{TEE}_j
\]

Let us assume we have \( k \) node runners (each with a TEE), but only \( l \) of them need secure communication (\( l < k \)) due to the sharding,
The process involves establishing secure communication channels among the required TEEs through a series of steps to ensure encrypted data transmission and mutual trust. 
This approach optimizes communication and enhances security without unnecessary complexity.

Note that these TEEs are both attested and verified, so they can establish secure communication channels using a key exchange protocol. 
This can be done using the Diffie-Hellman key exchange \cite{merkle1978secure}, where each pair of TEEs generates a shared secret key:

\[
\text{Shared Secret Key}_{ij} = g^{a_ib_j} \mod p.
\]

From this shared secret key, a session key is derived using a key derivation function (KDF):

\[
\text{Session Key}_{ij} = \text{KDF}(\text{Shared Secret Key}_{ij}).
\]

The session key is used to encrypt the data transmitted between the TEEs, ensuring confidentiality and integrity:

\[
\text{Encrypted Data}_{ij} = \text{AES}_{\text{Session Key}_{ij}}(\text{Data}).
\]

By following these steps, we ensure that the TEEs needing secure communication can transmit data securely and efficiently. This method reduces the reliance on the orchestrator for data routing, minimizes latency, and leverages the computational strengths of both GPU TEEs and CPU TEEs. 
The secure communication channels maintain data privacy and integrity throughout the decentralized AI inference process.

\textbf{Innovation 2: Heterogeneous TEE Scheduling}.
In our system, TEEs are heterogeneous, meaning they vary in type and computational power, with some nodes supporting GPU TEEs and others only CPU TEEs. To manage this diversity, we employ a dynamic scheduling strategy based on initial attestation results. This attestation determines whether a node supports GPU TEE or CPU TEE only. The motivation behind this is to leverage the strengths of each type of TEE, ensuring that computationally intensive tasks are handled by more capable nodes, thereby optimizing overall system performance.

\[
\text{TEE Type} = \begin{cases} 
\text{GPU TEE}, & \text{if GPU TEE requirements met}; \\
\text{CPU TEE}, & \text{if only CPU TEE requirements met}.
\end{cases}
\]

The dynamic scheduling strategy involves several innovative steps:
\begin{enumerate}
    \item \textbf{Initial Attestation}: This step determines the capabilities of each node, identifying GPU TEE or CPU TEE support. By understanding the hardware capabilities, we can tailor the workload distribution to maximize efficiency and performance.
    \item \textbf{Resource Allocation}: Based on the attestation outcomes, different shards of the AI model \(M\) are assigned to different TEEs according to their computational capabilities. More computationally intensive tasks are allocated to nodes with GPU TEEs, while less demanding tasks are assigned to CPU TEEs. This approach ensures that each node operates within its optimal performance range, enhancing the overall efficiency of the decentralized system.
    \item \textbf{Execution and Communication}: Establishing encrypted channels for direct communication between TEEs with computational dependencies reduces the orchestrator's load and improves efficiency. This direct TEE-to-TEE communication is critical for maintaining low latency and high throughput in distributed AI inference tasks.
\end{enumerate}

\textbf{Innovative Motivations}.
The innovations in our use of TEEs are driven by several key motivations:
\begin{itemize}
    \item \textbf{Security and Privacy}: Ensuring the confidentiality and integrity of user data and AI models is paramount. By using TEEs and secure communication channels, we guarantee that sensitive information is protected throughout the computation process.
    \item \textbf{Efficiency and Performance}: Optimizing the allocation of computational tasks based on the capabilities of heterogeneous TEEs ensures that the system operates efficiently. This dynamic scheduling reduces unnecessary delays and maximizes the use of available resources.
    \item \textbf{Scalability}: As the number of nodes increases, the system must efficiently manage communication and computation across a distributed network. Our innovations enable seamless scaling without compromising on performance or security.
\end{itemize}

By implementing these innovations, we effectively manage the decentralized AI inference process, ensuring secure, efficient, and optimal performance across heterogeneous TEEs. This approach leverages the strengths of both GPU and CPU TEEs, providing robust security measures and dynamic resource allocation to meet diverse computational needs.

\end{document}